\documentclass[aps,prl,superscriptaddress,twocolumn]{revtex4-2}

\usepackage{graphicx,amssymb,amsfonts,amsmath,chemarr,color,commath,hyperref}
\usepackage[version=4]{mhchem}
\usepackage{float}
\usepackage[dvipsnames]{xcolor}

\newcommand{\beginsupplement}{%
\setcounter{page}{1}
        \setcounter{equation}{0}
        \renewcommand{\theequation}{S\arabic{equation}}%
        \setcounter{table}{0}
        \renewcommand{\thetable}{S\arabic{table}}%
        \setcounter{figure}{0}
        \renewcommand{\thefigure}{S\arabic{figure}}%
     }
     \newcommand{\zstroke}{%
  \text{\ooalign{\hidewidth -\kern-.3em-\hidewidth\cr$z$\cr}}%
}

\begin{document}

\title{Multigenerational memory in bacterial size control}

\author{Motasem ElGamel}
\affiliation{Department of Physics and Astronomy, University of Pittsburgh, Pittsburgh, PA, 15260, USA}

\author{Harsh Vashistha}
\thanks{Present address: Department of Molecular, Cellular and Developmental Biology, Yale University, New Haven, CT, 06511, USA}
\affiliation{Department of Physics and Astronomy, University of Pittsburgh, Pittsburgh, PA, 15260, USA} 

\author{Hanna Salman}
\affiliation{Department of Physics and Astronomy, University of Pittsburgh, Pittsburgh, PA, 15260, USA}

\author{Andrew Mugler}
\email{andrew.mugler@pitt.edu}
\affiliation{Department of Physics and Astronomy, University of Pittsburgh, Pittsburgh, PA, 15260, USA}

\begin{abstract}
Cells maintain a stable size as they grow and divide. Inspired by the available experimental data, most proposed models for size homeostasis assume size control mechanisms that act on a timescale of one generation. Such mechanisms lead to short-lived autocorrelations in size fluctuations that decay within less than two generations. However, recent evidence from comparing sister lineages suggests that correlations in size fluctuations can persist for many generations. Here we develop a minimal model that explains these seemingly contradictory results. Our model proposes that different environments result in different control parameters, leading to distinct inheritance patterns. Multigenerational memory is revealed in constant environments but obscured when averaging over many different environments. Inferring the parameters of our model from {\it Escherichia coli} size data in microfluidic experiments, we recapitulate the observed statistics. Our work elucidates the impact of the environment on cell homeostasis and growth and division dynamics.
\end{abstract}

\maketitle
Cell size is a dynamic property of cells important for optimizing nutrient intake \cite{chien2012cell, turner2012cell}, accommodating intracellular content \cite{turner2012cell, marshall2012determines}, and maintaining uniformity in tissues \cite{ginzberg2015being}. Cell size fluctuations are significant, yet constrained \cite{wang2010robust}, suggesting active mechanisms of size control that go beyond initiating division a certain amount of time after birth \cite{amir2014cell, willis2017sizing, facchetti2017controlling, ho2018modeling}. Experiments and theory in recent years have revealed different phenomenological classes of size control \cite{amir2014cell, taheri2015cell, soifer2016single, susman2018individuality, facchetti2019reassessment}; connections between control of size, growth, and DNA replication \cite{amir2014cell, ho2015simultaneous, barber2017details, si2019mechanistic, witz2019initiation, berger2022robust}; and a surprising degree of heterogeneity in control mechanisms across, and even within, species \cite{tanouchi2015noisy, willis2017sizing, susman2018individuality}. Despite tremendous progress, basic questions remain open. In particular, it is still unclear whether deviations from the average size dissipate over one or many generations, and why the measured control parameters appear to vary so widely, even within lineages of the same population \cite{susman2018individuality}.

\begin{figure}
\centering
\includegraphics[width=1\linewidth]{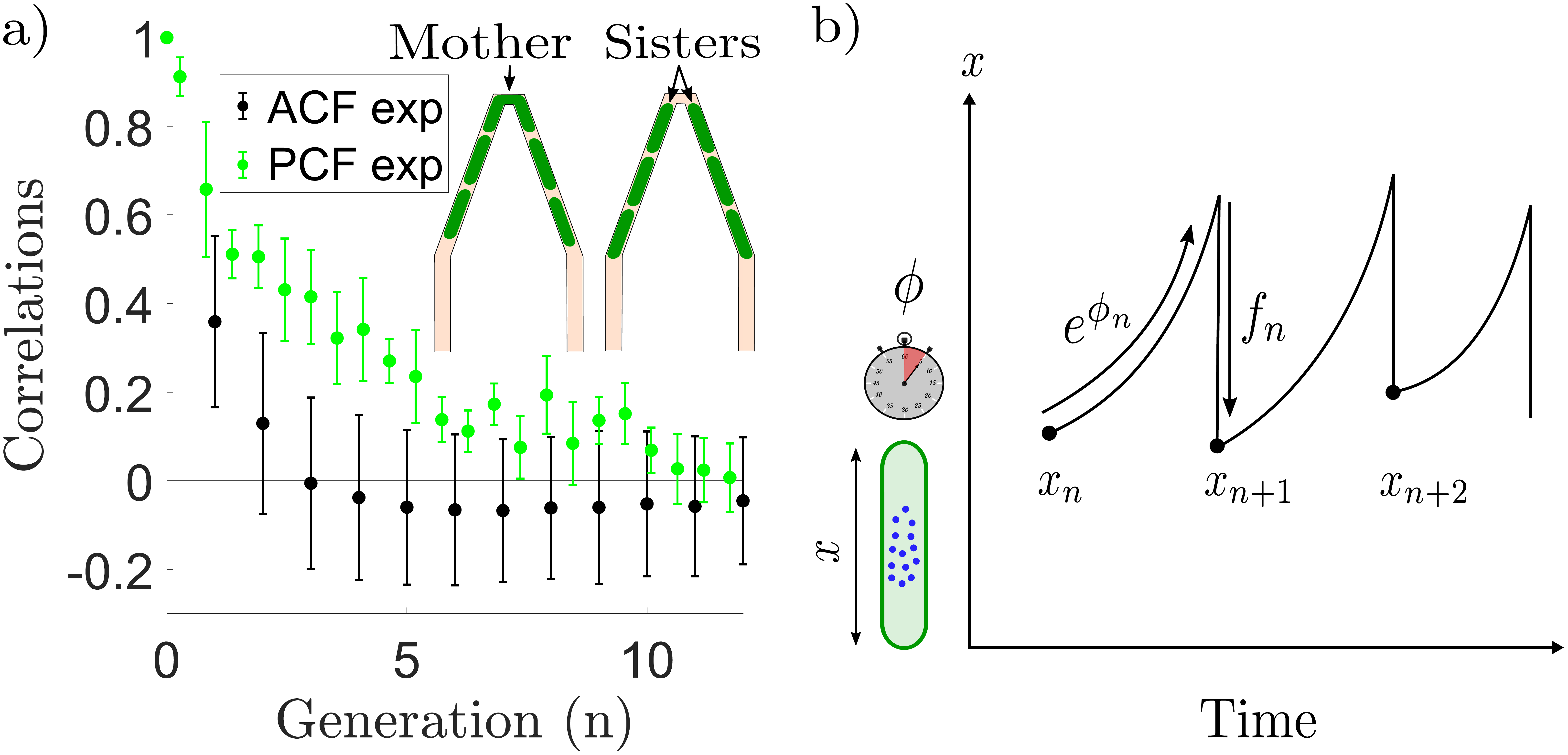}
\caption{(a) In the `sisters machine' (inset), a mother cell initiates two sister lineages in a common V-shaped channel \cite{harsh2021non}. Experiments \cite{harsh2021non} show that the autocorrelation function (ACF, black) for the cell birth size decays more quickly than the Pearson cross-correlation function between sister lineages (PCF, green). Here $n=0$ is the shared mother cell. (b) A cell grows exponentially from an initial birth size $x_n$ to a final division size $x_n e^{\phi_n}$, then divides by a fraction $f_n$.}
\label{fig1}
\end{figure}

Most experiments suggest that deviations from the average cell size last for only a generation or so. Specifically, microfluidic experiments with bacteria using devices such as the ``mother machine'' \cite{wang2010robust} generally find an exponentially decaying autocorrelation function (ACF) in cell birth size $A_n=e^{-n/n_A}$ with $n_A\approx1$ generation \cite{taheri2015cell, tanouchi2015noisy, susman2018individuality, si2019mechanistic}. Recently, however, experiments that track two lineages born from the same mother cell (a ``sisters machine'') \cite{harsh2021non} have found something different. Measuring the Pearson cross-correlation function (PCF) between birth sizes in these experiments has also revealed an exponential decay, $P_n = e^{-n/n_P}$, but with $n_P \approx 3.5$ generations (Fig.\ \ref{fig1}a, green). Surprisingly, these same experiments show $n_A\approx1$ generation for the lineages' ACF (Fig.\ \ref{fig1}a, black), consistent with the mother machine experiments (see Supplemental Material \cite{supp} for details of how correlation functions are calculated in theory and experiments). This raises the question of whether size deviations last for only a generation, as implied by $n_A$, or for multiple generations, as implied by $n_P$. More generally, it raises the question of how a signal is transiently more correlated with another signal than with itself.

It is expected that size deviations dissipate within a generation in the context of ``adder'' size control \cite{campos2014constant, soifer2016single, harris2016relative, willis2017sizing, ho2018modeling, si2019mechanistic}. Adder control means that a cell adds a constant amount to its birth size before dividing. To see the connection between adder control and how long size deviations last, consider a cell born with size $x_n$ that grows exponentially for an elapsed phase $\phi_n$, the product of the growth rate and cell cycle time (Fig.\ \ref{fig1}b). After division, the daughter with fraction $f_n$ will have birth size $x_{n+1} = f_nx_ne^{\phi_n}$. Defining $\epsilon_n=\ln(x_n/x_*)$ as the logarithmic deviation of the cell's birth size from the population-averaged birth size $x_*$, this expression becomes
\begin{equation}
\label{epsilon}
    \epsilon_{n+1}=\epsilon_n+\delta_n+\eta_n,
\end{equation}
where $\delta_n=\phi_n-\ln 2$ and $\eta_n=\ln (2f_n)$ are deviations of the phase and fraction from their expected values for size doubling. Experiments in {\it Escherichia coli} have shown that $\eta_n$ is Gaussian and uncorrelated between generations \cite{KOHRAM2021955}. In this case, size control implies that the phase corrects for deviations in the birth size \cite{osella2014concerted, campos2014constant, taheri2015cell, tanouchi2015noisy, susman2018individuality},
\begin{equation}
\label{delta1}
    \delta_n=-\beta\epsilon_n+\xi_n,
\end{equation}
where the homeostasis parameter $\beta$ sets the strength of the correction, and $\xi_n$ is uncorrelated Gaussian noise in the correction process. The values $\beta = 0$, $1/2$, and $1$ correspond to the ``timer'', adder, and ``sizer'' rules, respectively \cite{amir2014cell, susman2018individuality}. Experiments in bacteria generally observe a range of $\beta$ values, centered around $1/2$ corresponding to the adder rule \cite{tanouchi2015noisy, willis2017sizing, jun2018fundamental, susman2018individuality}.

Combining Eqs.\ \ref{epsilon} and \ref{delta1} gives a process $\epsilon_{n+1} = (1-\beta)\epsilon_n + \eta_n + \xi_n$ whose ACF $A_n = (1-\beta)^n$ and PCF $P_n = (1-\beta)^{2n}$ are straightforward to calculate \cite{supp}. For $\beta = 1/2$, we thus have $n_A = -1/\ln(1-\beta)\approx1.4$ generations and $n_P=n_A/2\approx0.7$ generations. While we see that $n_A$ is about one generation for adder control, this framework cannot explain why $n_P$ is observed in experiments to be multiple generations. Indeed, it is not clear from this framework how $n_P$ could be larger than, not smaller than, $n_A$. Instead, we see that two noisy signals decorrelate twice as quickly from each other as each does from itself.

Here we resolve this disagreement between theory and experiment by going beyond the standard model of cell size control in Eqs.\ \ref{epsilon} and \ref{delta1}. Our fundamental premise is that the environment plays a defining role in setting the size control parameters, and that different channels within a microfluidic device are subject to different environments \cite{yang2018analysis, stawsky2022multiple}. Because obtaining correlation functions from data often requires averaging over many channels to obtain sufficient statistics \cite{tanouchi2015noisy, susman2018individuality, harsh2021non}, we hypothesize that the averaging process obscures long timescales in some correlation functions ($A_n$) but not others ($P_n$). Inferring the parameters of our model from single-lineage autocorrelation data in {\it E.\ coli}, we find that this is indeed the case, suggesting that size correlations are multigenerational but dynamically diverse across environments. Our results suggest that size autocorrelations are compatible with an adder rule on average, but reveal the strong influence of a heterogeneous environment on individual lineages.

Before describing our main model, we first rule out the possibility that short autocorrelations and long cross-correlations between two lineages can be explained by the presence of common environmental fluctuations \cite{elowitz2002stochastic}. In principle, a signal with long intrinsic memory would exhibit short autocorrelations if this memory were overpowered by short-lived environmental noise. If this noise were common to both signals, the long memory would be expected to survive in the signal difference and therefore in the cross-correlation function. To investigate this possibility, we replace the noise term $\xi_n$ in Eq.\ \ref{delta1} with a long-lived, lineage-intrinsic component $y_n^{(i)}$ and a short-lived, environmental component $\chi_n$,
\begin{equation}
\label{delta2}
    \delta_n^{(i)}=-\beta \epsilon_n^{(i)} + y_n^{(i)} + \chi_n,
\end{equation}
where $i=\{1, 2\}$ denotes each of the two sister lineages. Note that the intrinsic component $y_n^{(i)}$ depends on the lineage $i$, whereas the environmental component $\chi_n$ does not. By giving the intrinsic component the dynamics
$y_{n+1}^{(i)} = \mu y_n^{(i)} + \zeta_n^{(i)}$
with uncorrelated Gaussian noise $\zeta_n^{(i)}$, we allow for long-lived memory that approaches $(1-\mu)^{-1}$ generations as $\mu$ approaches one. A natural interpretation of $y$ is the fluctuations in cellular protein content that regulates a cell's growth and metabolism and is inherited from one generation to the next. The environmental component $\chi_n$ is uncorrelated Gaussian noise.

\begin{figure}[b]
\centering
\includegraphics[width=1\linewidth]{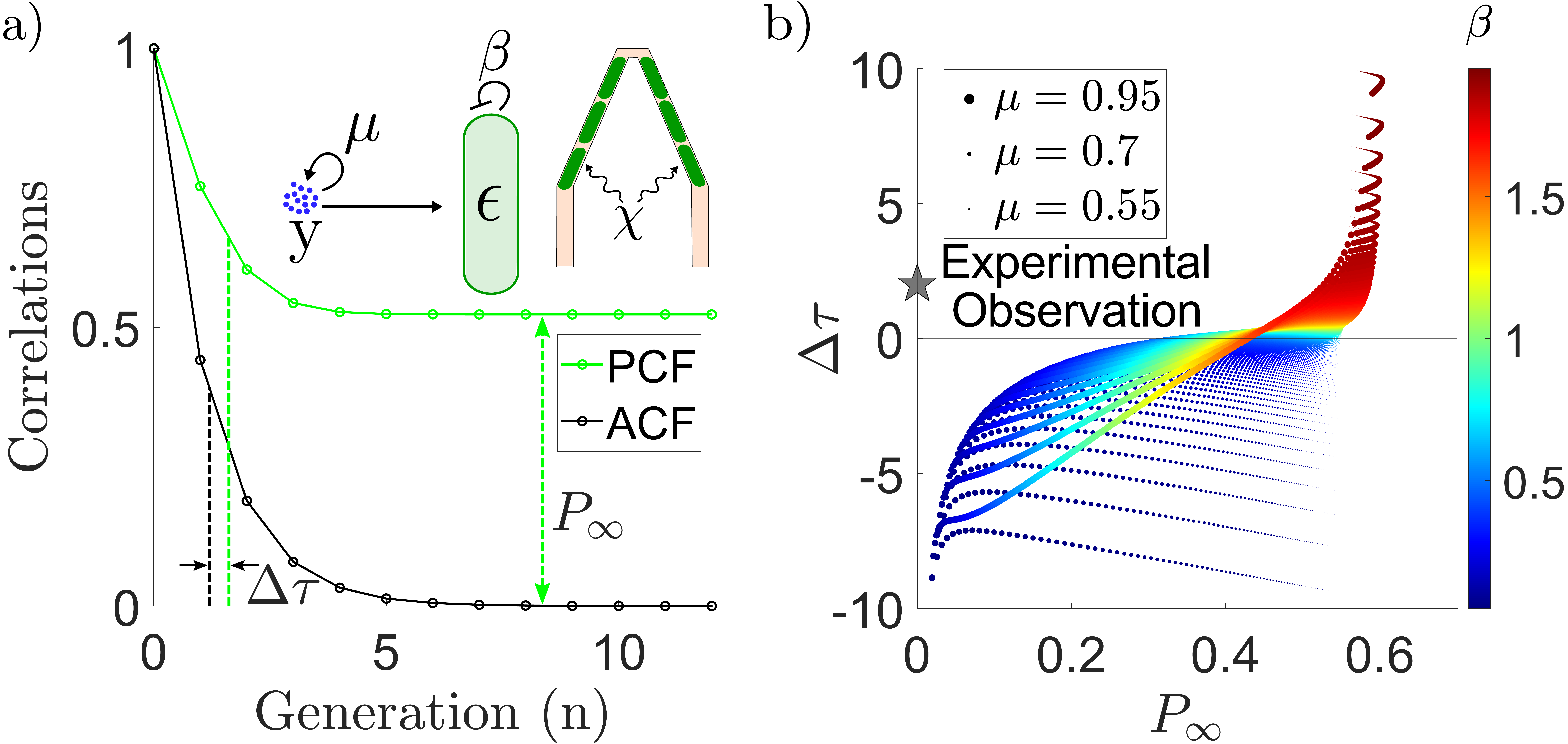}
\caption{The presence of common environmental fluctuations cannot explain the short-lived ACF and long-lived PCF observed in experiments (Fig.\ \ref{fig1}a). (a) A long-lived protein $y$ regulates cell size in two lineages subject to fast environmental noise $\chi$ (inset). The PCF (Eq.\ \ref{P1}) decays more slowly than the ACF (Eq.\ \ref{A1}) but has a large asymptote $P_\infty$. (b) No parameters can explain the experimental timescale difference and zero asymptote. $\beta=0.6$ and $\mu=0.3$ in a; $\sigma^2_\eta=0.0225$, $\sigma^2_\zeta=0.01$ and $\sigma^2_\chi= 0.04$ in a and b.}
\label{fig2}
\end{figure}

\begin{figure*}
\centering
\includegraphics[width=.95\linewidth]{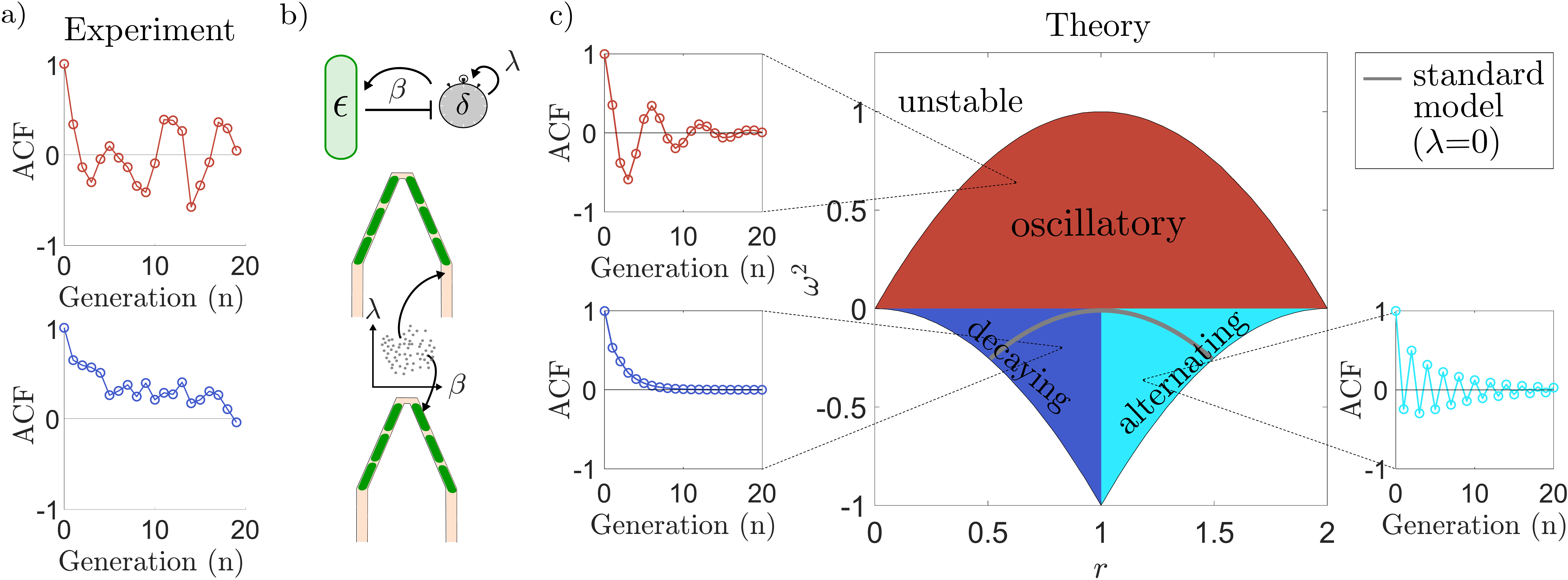}
\caption{Different dynamics in different environments. (a) Experimental birth size ACFs from single lineages in different channels \cite{harsh2021non} ranges from oscillatory to simple decay. (b) Our model includes size correction ($\beta$) and phase dependence ($\lambda$), with $\beta$ and $\lambda$ unique for each channel. (c) The model exhibits three stable dynamic regimes for the ACF depending on the values of the damping rate $r = (1+\beta-\lambda)/2$ and the squared underdamped frequency $\omega^2 = \beta-r^2$.}
\label{fig3}
\end{figure*}

Eliminating $\delta_n^{(i)}$ from Eqs.\ \ref{epsilon} and \ref{delta2} gives $\epsilon_{n+1}^{(i)} = (1-\beta) \epsilon_{n}^{(i)} + y_{n}^{(i)} + \eta_n^{(i)} + \chi_n$, a dynamics for size fluctuations $\epsilon$ that depends on $\epsilon$ itself and on $y$, as depicted in the inset of Fig.\ \ref{fig2}a. We solve for the ACF and PCF of $\epsilon_n^{(i)}$ by explicit iteration \cite{supp}. Defining $b=1-\beta$, we obtain
\begin{align}
\label{A1}
    A_n &\propto c_1 b^n + c_2 \mu^n, \\
\label{P1}
    P_n &\propto c_0 + c_3 b^{2n} + c_4 \mu^{2n} + c_5 b^n\mu^n,
\end{align}
where $c_0=\sigma^2_\chi/f$, $c_1=(\sigma^2_\chi+\sigma^2_\eta)/f-b\sigma^2_\zeta/fgh$, $c_2=\mu\sigma^2_\zeta/ghk$, $c_3=\sigma^2_\zeta/fg^2+\sigma^2_\eta/f$, $c_4=\sigma^2_\zeta/g^2k$, $c_5=-2\sigma^2_\zeta/g^2h$, $f=1-b^2$, $g=\mu-b$, $h=1-\mu b$, and $k=1-\mu^2$. Each $\sigma^2_i$ is the variance of the corresponding noise term, and the proportionality constants in Eqs.\ \ref{A1} and \ref{P1} are set by the normalization condition $A_0=P_0=1$. Because Eqs.\ \ref{A1} and \ref{P1} are not single exponential decays, we define a characteristic timescale as \cite{fancher2020diffusion} $\tau_C = \sum_{n=0}^\infty (C_n-C_\infty)/(C_0-C_\infty)$ for $C\in\{A,P\}$. Neglecting the fraction noise $\sigma^2_\eta$ (experiments show that $\sigma_\eta \approx 10\%$ \cite{susman2018individuality}), this gives
\begin{align}
\label{Pinf}
P_\infty &= \frac{\rho}{\rho+\ell},\\
\label{tauA}
\tau_A &= \frac{1}{\beta} + \left(\frac{1}{1-\mu}\right)\left[\frac{\mu(1+b)}{1+\mu b+\rho}\right],\\
\label{tauP}
\tau_P &= \frac{1}{1-b^2}+\frac{1}{1-\mu^2}+\frac{2b\mu}{1-b^2\mu^2}.
\end{align}
where $\rho=\sigma^2_\chi hk/\sigma^2_\zeta$ and $\ell=(hk+fh-2fk)/g^2$. We see that as $\mu$ approaches one, the second term in Eq.\ \ref{tauP} dominates, and the PCF indeed becomes long-lived. However, in order for the ACF to remain short-lived ($\tau_A\sim 1/\beta$), we see that the second term in Eq.\ \ref{tauA} must remain small, requiring $\rho\gg1$. This condition increases the long-time cross-correlation $P_\infty$, as seen in Eq.\ \ref{Pinf}.

The requirement $\rho=\sigma^2_\chi hk/\sigma^2_\zeta\gg1$ makes sense because in order for short extrinsic noise to wash out long intrinsic memory, the noise must be strong ($\sigma^2_\chi\gg\sigma^2_\zeta$). The fact that this then increases $P_\infty$ also makes sense because strong extrinsic noise leaves two signals strongly correlated indefinitely. The net result is that the PCF timescale cannot be longer than the ACF timescale without a large PCF asymptote, as illustrated in Fig.\ \ref{fig2}a. In fact, numerically probing all values of $\mu$ and $\beta$ (with $\sigma^2_\eta$ nonzero), we find no value of $\Delta\tau = \tau_P-\tau_A$ and $P_\infty$ consistent with the experimental observations of $\Delta\tau>0$ and $P_\infty=0$ (Fig.\ \ref{fig2}b). We conclude that the observations in Fig.\ \ref{fig1}a cannot be explained by the presence of common environmental fluctuations.

If the environment is not providing strong fluctuations, is it playing an alternative role? To obtain insight into this question, we recognize that ACFs from individual lineages in different channels exhibit different dynamic behaviors, ranging from simple decay to oscillations (Fig.\ \ref{fig3}a) \cite{tanouchi2015noisy, susman2018individuality, harsh2021non}.
Oscillations in single-lineage ACFs could reflect insufficient data \cite{tanouchi2015noisy, si2019mechanistic}, although they persist even for lineages with hundreds of generations \cite{susman2018individuality}, suggesting that they may reflect genuine overcorrection in size control. We have checked using simulations that genuine oscillations are detectable for $N \ge 20$ generations (Fig.\ S1 \cite{supp}), and therefore we only analyze experimental lineages at least this long.

To explain the heterogeneity of dynamic behaviors, we hypothesize that the size control parameters are a function of the environment, and that different channels have different environments. Environmental heterogeneity could be due to nutrient gradients on the lengthscale of the entire microfluidic device or mechanical differences among channels (mechanical forces limit growth in narrow channels, and actual channel widths can be different from designed widths \cite{yang2018analysis}). Indeed, recent experimental analysis has shown that cells in different channels fluctuate around different homeostatic set points \cite{KOHRAM2021955, stawsky2022multiple}, consistent with the hypothesis of different environments.

To investigate this hypothesis, we modify Eq. \ref{delta1} as
\begin{align}
\label{delta3}
    \delta_n^{(i)}=-\beta \epsilon_n^{(i)} + \lambda \delta_{n-1}^{(i)} + \xi_n^{(i)},
\end{align}
where now $\xi_n^{(i)}$ is uncorrelated Gaussian noise. The new term in Eq.\ \ref{delta3}, $\lambda \delta_{n-1}^{(i)}$, introduces a dependence of the phase on its value in the previous generation.
Physiologically, this dependence could result from the inheritance of fluctuations in key growth control factors, such as ribosomes, RNA polymerases, and other proteins, from one generation to the next \cite{KOHRAM2021955}. The dependence could be positive or negative, depending on whether the inherited factor primarily affects the growth rate or the cell cycle time \footnote{Inheritance of protein content could affect the growth rate and the cell cycle time in different ways. Inheritance of growth proteins (e.g., ribosomes) could result in positive $\lambda$ \cite{KOHRAM2021955}: a mother with more-than-average growth proteins will grow fast and produce a daughter with more-than-average growth proteins that also grows fast. Inheritance of division proteins (e.g., FtsZ) that may need to reach a fixed threshold at division could result in negative $\lambda$: a mother born with more-than-average division proteins will take less time to reach the threshold and divide, but if the threshold is met precisely, the daughter will then be born with a closer-to-average amount of division proteins, and thus take an average amount of time to divide. This will correct the mother’s deviation in the cell cycle time, corresponding to a negative dependence.}. Indeed, a similar term emerges naturally (along with $\beta$) from a systematic autoregression analysis of single-cell growth data \cite{KOHRAM2021955}, providing experimental evidence for the dependence \footnote{In \cite{KOHRAM2021955}, the phase variable depends on the growth rate, which depends on its value in the previous generation. In our model, the phase variable depends directly on its value in the previous generation. Both models fall in the same phenomenology class (capable of oscillations and decay), and because we seek the simplest member of this class, we choose the simpler model.}. In principle, $\lambda$ could be perturbed by modulating the growth control factors, and $\beta$ is known to increase for slower-growing cells \cite{wallden2016synchronization, si2019mechanistic}.

As illustrated in Fig.\ \ref{fig3}b (top), Eqs.\ \ref{epsilon} and \ref{delta3} contain feedback via (i) $\beta$, which compensates for a larger birth size via a smaller phase, and (ii) $\lambda$, which accounts for the generational dependence of the phase. Together, these terms produce damped, oscillatory dynamics, which can be seen by the following mapping: rearranging Eqs.\ \ref{epsilon} and \ref{delta3} as $\epsilon_{n+1} -\epsilon_n = \delta_n + \eta_n$ and $\delta_{n+1} - \delta_n = -\beta\epsilon_n + (\lambda-\beta-1)\delta_n + (\xi_{n+1} - \beta\eta_n)$, we can approximate their lefthand sides as time derivatives and combine them, yielding $\ddot{\epsilon} + 2r\dot{\epsilon} + (r^2+\omega^2)\epsilon = \psi$, where $r = (1+\beta-\lambda)/2$, $\omega^2 = \beta-r^2$, and $\psi_n = \xi_{n+1} + \eta_{n+1} - \lambda\eta_n$. These are the dynamics of a simple harmonic oscillator with damping rate $r$ and underdamped frequency $\omega$, driven by noise $\psi$.

Importantly, the parameters $\beta$ and $\lambda$ (and thus $r$ and $\omega^2$) are the same for each of the two lineages $i$ in a channel but vary from channel to channel (Fig.\ \ref{fig3}b, bottom). The question is whether different values of $r$ and $\omega^2$ can capture the dynamic heterogeneity observed in the experiments, and whether averaging over these values results in the observed auto- and cross-correlations.

To address this question, we solve for the ACF and PCF of $\epsilon_n^{(i)}$ in Eqs.\ \ref{epsilon} and \ref{delta3} using the Z-transform (the discrete-time analog of the Laplace transform) \cite{supp}. We obtain
\begin{align}
\label{A2}
    A_n &\propto q_- a_-^n + q_+ a_+^n,\\
\label{P2}
    P_n &\propto s_- a_-^{2n} + s_+ a_+^{2n} - 2s a_-^n a_+^n, 
\end{align}
where $a_{\pm} = 1-r \pm \sqrt{-\omega^2}$, the coefficients $q_\pm$, $s_\pm$, and $s$ are functions of $a_\pm$ and the noise strengths $\sigma_\eta^2$ and $\sigma_\xi^2$ \cite{supp}, and again the proportionality constants are set by $A_0=P_0=1$. Stability requires $|a_\pm|<1$, equivalent to the conditions $\omega^2>-r^2$, $\omega^2>-(r-2)^2$, and $\omega^2<-r(r-2)$ \cite{supp} (bordering parabolas in Fig.\ \ref{fig3}c). Damped oscillations occur when $\omega^2>0$ (horizontal line in Fig.\ \ref{fig3}c). Alternation, which is unique to discrete systems when the autocorrelation function is dominated by a term with a factor of $(-1)^n$, occurs when $r>1$ \cite{supp}, (vertical line in Fig.\ \ref{fig3}c). Together these conditions give three dynamic regimes, illustrated in Fig.\ \ref{fig3}c (see Fig.\ S2a \cite{supp} for these regimes in the space of $\beta$ and $\lambda$). In particular, we see that simple decay (blue) and oscillations (red) are possible, as observed in the data. Oscillations are not possible in the standard model with $\lambda = 0$ (gray parabola in Fig.\ \ref{fig3}c).

\begin{figure}
\centering
\includegraphics[width=1\linewidth]{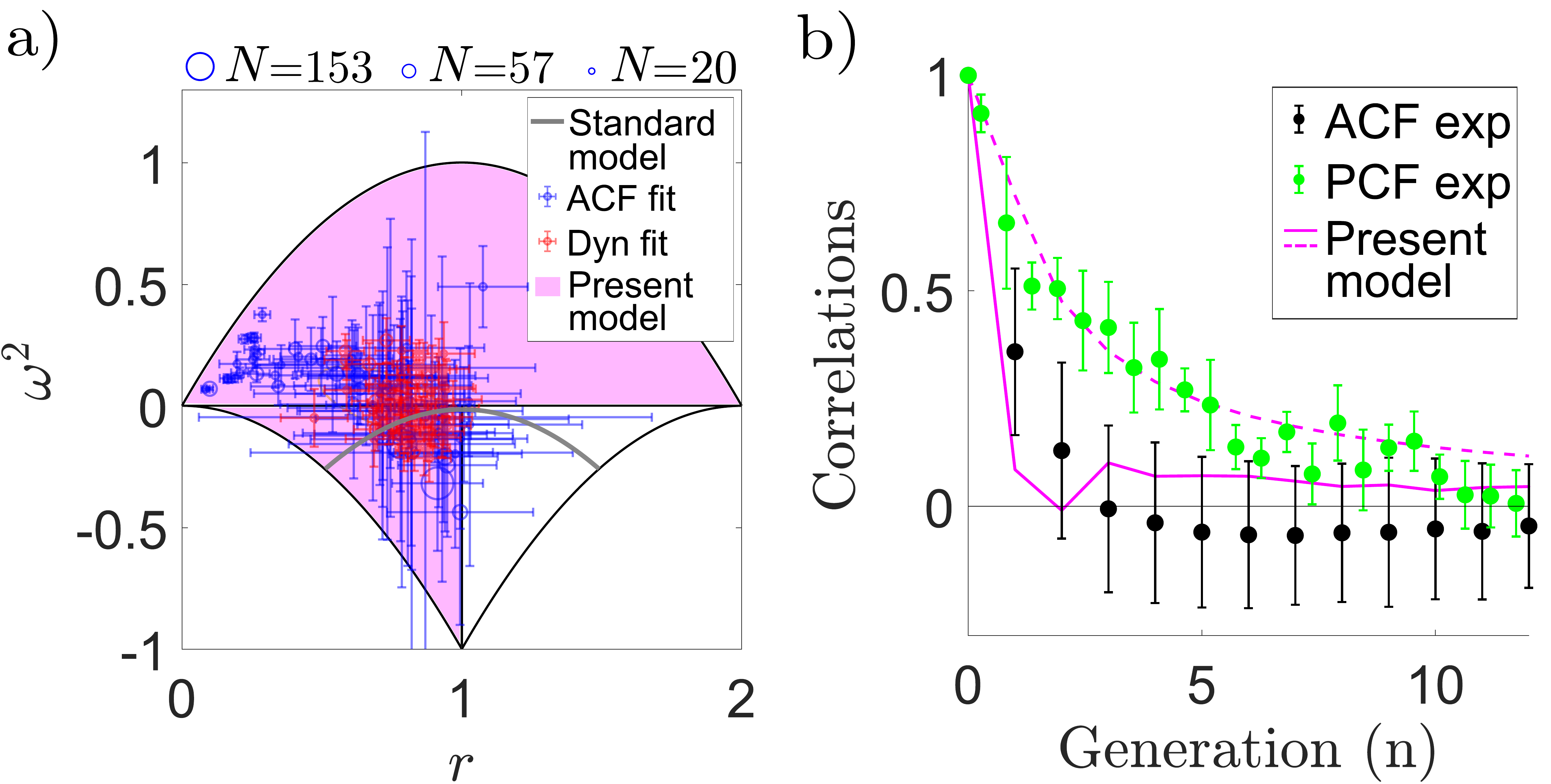}
\caption{Comparing theory and experiment. (a) $r$ and $\omega^2$ fitted from experimental single-lineage dynamics (red, Eq.\ \ref{delta3}) or ACFs (blue, Eq.\ \ref{A2}). Circle size: lineage length $N$ (top legend; see Fig.\ S2b for clarity). (b) ACF (Eq.\ \ref{A2}) and PCF (Eq.\ \ref{P2}), averaged over $r$ and $\omega^2$ values from a (pink regions), compared with experimental results. $\sigma^2_\eta =0.09$ and $\sigma_\xi^2 =0.04$ in a and b.}
\label{fig4}
\end{figure}

Addressing whether our model explains the correlation data requires determining in which dynamic regimes the experiments lie. To this end, we estimate the parameters $r$ and $\omega^2$ in two ways. First, we perform a least-squares fit of Eq.\ \ref{delta3} to each single-lineage dynamics, as a planar equation for $\delta_n^{(i)}$ vs.\  ($\epsilon_n^{(i)}$, $\delta_{n-1}^{(i)}$); second, we fit Eq.\ \ref{A2} to each single-lineage ACF (see \cite{supp} for details). Consistent with our hypothesis (Fig.\ \ref{fig3}b, bottom), in both cases we allow the $r$ and $\omega^2$ values to be different for different channels, but we require them to be the same for sister lineages in the same channel by combining the two sums of squares during fitting (relaxing this constraint results in a similar distribution of fitted values, Fig.\ S2b \cite{supp}). The resulting values of $r$ and $\omega^2$ for the two methods are shown in Fig.\ \ref{fig4}a (red and blue, respectively; see Fig.\ S2a \cite{supp} for these data in the space of $\beta$ and $\lambda$). Values from data in a different growth condition also lie in the same parameter region (Fig.\ S2b \cite{supp}).

We see in Fig.\ \ref{fig4}a that the parameters inferred using either method generally lie in the decaying and oscillatory regimes (pink) but not the alternating regime (white). The parameters inferred from the ACF fits (blue) span a larger range than those inferred from the dynamics (red), but each case populates both regimes with various frequencies and damping strengths. Neither is confined to the standard model (gray parabola). Furthermore, the regime does not correlate with the length of the lineage (size of circle; see Fig.\ S2b for clarity), suggesting that observed dynamic features are not artifacts of insufficient data.

We therefore ask whether averaging over the decaying and oscillatory regimes (pink in Fig.\ \ref{fig4}a) is sufficient to explain the correlations observed in experiments. Performing this average, we obtain the results in Fig.\ \ref{fig4}b (pink). We see that the averaged ACF is a relatively smooth function that decays in about a generation, consistent with the experimental averaged ACF (Fig.\ \ref{fig4}b, black). Evidently, oscillations with different frequencies are largely washed out in the averaging process \cite{tanouchi2015noisy, susman2018individuality}, producing an apparent fast decay. We also see that the averaged PCF exhibits a longer timescale than the ACF, consistent with the experimental data (Fig.\ \ref{fig4}b, green). The reason is that an individual channel's PCF does not oscillate, even when the ACF does, because the PCF reports on the difference between two oscillatory signals, not the signals themselves. Consequently, the averaged PCF is sensitive to its longer-lived samples, whereas the averaged ACF appears short-lived due to the washout of many oscillation periods.

We have put forward a minimal model for cell size control that resolves the empirical paradox of short-lived autocorrelations but long-lived cross-correlations in cell size. We have found that cell size memory is longer-lived than previously appreciated but is obscured in autocorrelations due to destructive interference among many oscillation periods.
The results suggest that control parameters depend sensitively on the environment and that the environment varies considerably within a multi-channel microfluidic device, as has been suggested \cite{yang2018analysis} and demonstrated \cite{KOHRAM2021955, stawsky2022multiple} previously.

The dynamics in Eq.\ \ref{delta3} go beyond the standard model of cell size control ($\lambda=0$) and have a structure motivated by recent single-cell growth experiments \cite{KOHRAM2021955}. Parameters inferred from these dynamics (Fig.\ \ref{fig4}a, red) occupy the same regimes but nevertheless a narrower range than parameters inferred from the autocorrelation functions (Fig.\ \ref{fig4}a, blue), suggesting that the dynamics may be incomplete. A more accurate dynamical model might include nonlinear terms, relate more than two consecutive generations, or involve time-dependent parameters (we have checked in Fig.\ S3 \cite{supp} that temporal parameter fluctuations have little effect on the correlation functions). Eqs.\ \ref{epsilon} and \ref{delta3} are not unique in generating Eqs.\ \ref{A2} and \ref{P2}, and it will be interesting to see whether the dynamic control mechanism can be better pinpointed in future work.

Our central prediction that heterogeneous environments obscure multigenerational timescales in the averaged autocorrelation function could be tested by modulating the degree of heterogeneity in the channel environments. To the extent that the heterogeneity is nutrient-limited, it could be modulated either by flowing nutrients overtop the cell traps, which would reduce heterogeneity, or by inducing chemical gradients along the device, which would increase heterogeneity. Both are feasible options for future experiments.

We thank Michael Vennettilli for introducing us to the Z-transform. This work was supported by National Science Foundation grants MCB-1936761 and PHY-1945018 to A.M.\ and PHY-2014116 to H.S.


\providecommand{\noopsort}[1]{}\providecommand{\singleletter}[1]{#1}%
%

\clearpage

\onecolumngrid
\appendix

\beginsupplement

\begin{center}
    {\bf SUPPLEMENTAL MATERIAL}
\end{center}

\section{Equivalence of theoretical calculation and experimental computation of ACF and PCF}

For the ACF, we define the average on a per-lineage basis (Fig.\ \ref{fig3}c). Then, we average the ACFs over all lineages (Fig.\ \ref{fig4}b). This is standard practice in the analysis of experimental data \cite{tanouchi2015noisy, susman2018individuality, harsh2021non}, including the data in Figs.\ \ref{fig3}a (single lineage) and \ref{fig1}a/\ref{fig4}b (averaged ACFs). For the PCF, we take the average per pair of lineages, then we average the PCFs over all pairs (Fig.\ \ref{fig4}b). Experimentally, it is not possible to take the average per pair of lineages because there is only one pair per trap per generation. Therefore, experimentally, the global average is taken \cite{harsh2021non}, including in Fig.\ \ref{fig1}a/\ref{fig4}b. We now show that these two procedures are equivalent when lineage data is properly normalized with this lineage standard deviation as done in experiments. The definition of the global average used in experiments for the PCF is \cite{harsh2021non}
\begin{equation}
\label{pcf1}
P(t)=\frac{1}{\sigma_{y^{(1)}(t)} \sigma_{y^{(2)}(t)}}\sum^{N}_{i=1}{(y_i^{(1)}(t)-\langle y^{(1)}(t)\rangle) (y_i^{(2)}(t) -\langle y^{(2)}(t)\rangle)},
\end{equation}
where $y(t)$ is any variable of interest, $i$ indicates a pair of sisters (or equivalently, one trap of the sisters machine) and $N$ is the total number of pairs in the system. Since the sisters lineages are statistically identical, $\sigma_{y^{(1)}(t)}= \sigma_{y^{(2)}(t)}$ and we drop the superscript. This enables us to write eq.\ \ref{pcf1} as
\begin{equation}
\label{pcf2}
P(t)=\frac{1}{\sigma_{y(t)}^2}\sum^{N}_{i=1}{(y_i^{(1)}(t)-\langle y^{(1)}(t)\rangle) (y_i^{(2)}(t) -\langle y^{(2)}(t)\rangle)},
\end{equation}
Assuming there are $k$ sub-populations each containing $p$ number of pairs ($p_1, p_2, p_3, .., p_{k}$) such that $N=\sum^{k}_{\alpha}{p_\alpha}$, we can write eq.\ \ref{pcf1} as
\begin{equation}
\label{pcf3}
\begin{aligned}
P(t)&=\frac{1}{\sigma_{y(t)}^2}\big(\sum^{p_1}_{i=1}{(y_i^{(1)}(t)-\langle y^{(1)}(t)\rangle) (y_i^{(2)}(t) -\langle y^{(2)}(t)\rangle)}+\sum^{p_2}_{i=1}{(y_i^{(1)}(t)-\langle y^{(1)}(t)\rangle) (y_i^{(2)}(t) -\langle y^{(2)}(t)\rangle)}\\
&+..+\sum^{p_k}_{i=1}{(y_i^{(1)}(t)-\langle y^{(1)}(t)\rangle) (y_i^{(2)}(t) -\langle y^{(2)}(t)\rangle)}\big),
\end{aligned}
\end{equation}
which can be written more concisely as
\begin{equation}
\label{pcf4}
P(t)=\frac{1}{\sigma_{y(t)}^2}\sum^{k}_{\alpha=1}\sum^{p_\alpha}_{i_\alpha=1}{(y_{i_\alpha}^{(1)}(t)-\langle y^{(1)}(t)\rangle) (y_{i_\alpha}^{(2)}(t) -\langle y^{(2)}(t)\rangle)}.
\end{equation}
Similar argument can be used to show that $\sigma^2_y(t)$ can be written as $\sigma^2_y(t)=\sum^{k}_{\alpha=1}\sum^{p_\alpha}_{i_\alpha=1}{(y_{i_\alpha}^{(1)}(t)-\langle y^{(1)}(t)\rangle)}$. To write everything more concisely we make use of the observation that the numerator of eq.\ \ref{pcf4} is the sum of sub-populations covariances and the denominator is the sum of the sub-populations variances, giving
\begin{equation}
\label{pcf5}
P(t)=\frac{1}{\sum^{k}_{\alpha=1}{\sigma_{y_{\alpha}(t)}^2}}\sum^{k}_{\alpha=1}{cov(y_{\alpha}^{(1)}(t),y_{\alpha}^{(2)}(t))}.
\end{equation}
Normalizing each sub-population by the sub-population standard deviation so that each sub-population have a standard deviation of 1, making $cov(y_{\alpha}^{(1)}(t),y_{\alpha}^{(2)}(t))\rightarrow \frac{1}{\sigma^2_{y_\alpha}}cov(y_{\alpha}^{(1)}(t),y_{\alpha}^{(2)}(t))$ and $\sigma_{y_{\alpha}(t)}^2\rightarrow 1$. The theory assumes a very large number of lineages in the ensemble at steady state which makes the standard deviation of a sub-population at any moment of time equal to the total standard deviation of the sub-population ($\sigma_{y_{\alpha}(t)}^2=\sigma^2_{y_\alpha}$). Therefore, eq.\ \ref{pcf5} becomes
\begin{equation}
\label{pcf6}
P(t)=\frac{1}{k}\sum^{k}_{\alpha=1}{cov(y_{\alpha}^{(1)}(t),y_{\alpha}^{(2)}(t))/\sigma^2_{y_\alpha}},
\end{equation}
that is exactly an average of the sub-populations PCF's and can be written as
\begin{equation}
\label{pcf7}
P(t)=\frac{1}{k}\sum^{k}_{\alpha=1}{p_\alpha(t)}=\langle p(t) \rangle.
\end{equation}
In this picture, each sub-population is governed by the same dynamical parameters. Experimentally, we have only one lineage per a set of parameters. To make theoretical progress, we therefore consider each lineage as a sub-population of the larger ensemble.

\section{Standard model (Eqs.\ \ref{epsilon} and \ref{delta1})}
\noindent This model can be represented using one equation by substituting Eq. \ref{delta1} into Eq. \ref{epsilon}. This becomes
\begin{equation}
    \label{stmodel}
    \epsilon^{(i)}_{n+1}=(1-\beta)\epsilon^{(i)}_n + \eta^{(i)}_n +\xi^{(i)}_n.
\end{equation}
where the superscript $(i)$ indicates lineage and the subscript $n$ indicates generation. Iterating Eq. \ref{stmodel} explicitly we get

\begin{align*}
    \epsilon^{(i)}_{1}&= (1-\beta) \epsilon^{(i)}_{0} + \eta^{(i)}_{0} + \xi^{(i)}_0, \\
    \epsilon^{(i)}_{2}&= (1-\beta) \epsilon^{(i)}_{1} + \eta^{(i)}_{1} + \xi^{(i)}_1\\
    &=(1-\beta)^2\epsilon^{(i)}_{0}+(1-\beta)\eta^{(i)}_{0}+(1-\beta)\xi^{(i)}_{0}+ \eta^{(i)}_{1} + \xi^{(i)}_1,\\
    \epsilon^{(i)}_{3}&=(1-\beta)\epsilon^{(i)}_{2}+\eta^{(i)}_{2}+\xi^{(i)}_{2}\\
    &=(1-\beta)^3\epsilon^{(i)}_{0}+(1-\beta)^2\eta^{(i)}_{0}+(1-\beta)\eta^{(i)}_{1}+\eta^{(i)}_{2}+(1-\beta)^2\xi^{(i)}_{0}+(1-\beta)\xi^{(i)}_{1}+\xi^{(i)}_{2}.
\end{align*}
A pattern becomes evident and allows us to write a closed form for the cell size equation
\begin{equation}
\label{stdmodel2}
    \epsilon^{(i)}_{n}=(1-\beta)^n \epsilon^{(i)}_{0}+\sum^{n-1}_{m=0}{(\eta^{(i)}_{m}+\xi^{(i)}_{m})(1-\beta)^{n-m-1}},
\end{equation}
where $\epsilon^{(i)}_0$ is the initial cell size of lineage $i$.

\subsubsection{Autocorrelation Function}
   \noindent The size ACF is defined by
    \begin{equation}
    \label{auto1}
        A(n)\propto\lim_{n' \to \infty}{\langle \epsilon^{(i)}_{n+n'} \epsilon^{(i)}_{n'} \rangle}.
    \end{equation}
Substituting Eq. \ref{stdmodel2} into Eq. \ref{auto1} to calculate size ACF we find
\begin{align}
    A(n)&\propto \lim_{n' \to \infty}\big[(1-\beta)^{2n'+n} \langle\epsilon^{(i)}_{0}\epsilon^{(i)}_{0}\rangle+ \sum^{n+n'-1}_{m=0}\sum^{n'-1}_{k=0}{(\langle\eta^{(i)}_{m} \eta^{(i)}_{k}\rangle + \langle\xi^{(i)}_{m} \xi^{(i)}_{k}\rangle)(1-\beta)^{2n'+n-m-k-2}}\big],
    \end{align}
    where $\xi$ and $\eta$ are delta correlated white noises defined as
    \begin{equation}
    \label{noise0}
    \begin{aligned}
    \langle\xi^{(i)}_{m} \xi^{(j)}_{k}\rangle &= \sigma^2_{\xi}\delta_{mk}\delta_{ij},\\
        \langle\eta^{(i)}_{m} \eta^{(j)}_{k}\rangle &= \sigma^2_{\eta}\delta_{mk}\delta_{ij},
    \end{aligned}
        \end{equation}
and $\sigma^2_k$ is the variance of the corresponding noise. It can be read off from Eq. \ref{stdmodel2} that $1-\beta$ must be less than 1 for the model to be stable, allowing for the first term of the autocorrelation to go to zero in the limit $n' \to \infty$. The autocorrelation function then becomes
\begin{align}
    A(n)&\propto \lim_{n' \to \infty} \sum^{n'-1}_{k=0}{(\sigma^2_{\eta} + \sigma^2_{\xi})(1-\beta)^{2n'+n-2k-2}}.
    \end{align}
This summation is just a finite geometric series which can be done, and then we take the limit,
\begin{equation}
    A(n)\propto \frac{(1-\beta)^n}{\beta (2-\beta)},
\end{equation}
which when normalized becomes
\begin{equation}
    A(n)= (1-\beta)^n.
\end{equation}

\subsubsection{Pearson Correlation Function}
    \noindent The PCF is defined as
    \begin{align}
    \label{ps1}
    P(n)\propto\langle\epsilon^{(1)}_n\epsilon^{(2)}_n\rangle.
    \end{align}
Substituting Eq. \ref{stdmodel2} into \ref{ps1} results in
\begin{align}
    P(n)\propto
    (1-\beta)^{2n} \langle\epsilon^{2}_{0}\rangle,
    \end{align}
where $\epsilon_0^{(1)}$ is equal to $\epsilon_0^{(2)}$ due to the fact that sister cells share the same mother. Using Eqs. \ref{noise0} we could also eliminate noise correlation between sister cells. Normalizing the PCF we find
\begin{align}
    P(n)=
    (1-\beta)^{2n}.
    \end{align}

\section{Model with common environmental fluctuations (Eqs.\ \ref{epsilon} and \ref{delta2})}

\noindent Similar to the standard model, the same analysis can be repeated. We start by combining Eqs. \ref{epsilon} and \ref{delta2}, then iterate model equations to get
\begin{align*}
    \epsilon_{1}^{(i)}&= (1-\beta) \epsilon_{0}^{(i)} + y_{0}^{(i)} + \eta_0^{(i)} + \chi_{-1},\\
    \epsilon_{2}^{(i)}&= (1-\beta) \epsilon_{1}^{(i)} + y_{1}^{(i)} + \eta_1^{(i)} + \chi_0\\
    &=(1-\beta)^2\epsilon^{(i)}_{0}+(1-\beta)\chi_{-1}+\chi_{0}+(1-\beta)y^{(i)}_{0}+y^{(i)}_{1}+(1-\beta)\eta^{(i)}_{0}+\eta^{(i)}_{1},\\
    \epsilon^{(i)}_{3}&=(1-\beta)\epsilon^{(i)}_{2}+y^{(i)}_{2}+\eta^{(i)}_{2}+\chi_{1}\\
    &=(1-\beta)^3\epsilon^{(i)}_{0}+(1-\beta)^2\chi_{-1}+(1-\beta)\chi_{0}+\chi_{1}\\
    &\quad+(1-\beta)^2y^{(i)}_{0}+(1-\beta)y^{(i)}_{1}+y^{(i)}_{2}+(1-\beta)^2\eta^{(i)}_{0}+(1-\beta)\eta^{(i)}_{1}+\eta^{(i)}_{2}.
\end{align*}
Since $\chi$ is a delta correlated white noise, we can shift its index by $1$. We can find a closed form for the cell size
\begin{equation}
\label{s1}
    \epsilon^{(i)}_{n}=(1-\beta)^n \epsilon^{(i)}_{0}+\sum^{n-1}_{m=0}{(\chi_{m}+\eta^{(i)}_{m}+y^{(i)}_{m})(1-\beta)^{n-m-1}}.
\end{equation}
    A similar procedure can be done for the dynamics of $y$ (below Eq. \ref{delta2}),
\begin{equation}
y_{n+1}^{(i)} = \mu y_n^{(i)} + \zeta_n^{(i)}.
\end{equation}
The resulting closed form equation is
    \begin{equation}
    \label{s2}
        y^{(i)}_{n}=\mu^n y^{(i)}_{0} + \sum^{n-1}_{k=0}{\zeta^{(i)}_{k} \mu^{n-k-1}}.
    \end{equation}
    Eq. \ref{s2} can be inserted into Eq. \ref{s1} resulting in
    
    \begin{equation}
    \label{s3}
    \begin{aligned}
        \epsilon^{(i)}_{n}&=(1-\beta)^n \epsilon^{(i)}_{0}+\sum^{n-1}_{m=0}{(\chi_{m}+\eta^{(i)}_{m}+\mu^m y^{(i)}_{0} + \sum^{m-1}_{k=0}{\zeta^{(i)}_{k} \mu^{m-k-1}})(1-\beta)^{n-m-1}}\\
        &=(1-\beta)^n \epsilon^{(i)}_{0}+\sum^{n-1}_{m=0}{(\chi_{m}+\eta^{(i)}_{m}+\mu^m y^{(i)}_{0})(1-\beta)^{n-m-1}}+ \sum^{n-1}_{m=1}{\sum^{m-1}_{k=0}{\zeta^{(i)}_{k} \mu^{m-k-1}}(1-\beta)^{n-m-1}}.
    \end{aligned}
    \end{equation}
    
    \subsubsection{Autocorrelation Function}
    \noindent Substituting Eq. \ref{s3} into \ref{auto1} we get
    \begin{align*}
    A(n)&\propto \lim_{n' \to \infty}\big[(1-\beta)^{2n'+n} \langle\epsilon^{(i)}_{0}\epsilon^{(i)}_{0}\rangle+ \langle\epsilon^{(i)}_{0} y^{(i)}_{0}\rangle ( \sum^{n+n'-1}_{m=0}{\mu^m (1-\beta)^{2n'+n-m-1}} + \sum^{n'-1}_{m=0}{\mu^m (1-\beta)^{2n'+n-m-1}} )\\
    &+\langle y^{(i)}_{0} y^{(i)}_{0}\rangle\sum^{n+n'-1}_{m=0}\sum^{n'-1}_{k=0}{\mu^{m+k} (1-\beta)^{2n'+n-m-k-2} }+ \sum^{n+n'-1}_{m=0}\sum^{n'-1}_{k=0}{(\langle\chi_{m} \chi_{k}\rangle+\langle\eta^{(i)}_{m} \eta^{(i)}_{k}\rangle)(1-\beta)^{2n'+n-m-k-2}}\\&+\sum^{n'-1}_{m=1}\sum^{m-1}_{k=0}\sum^{n+n'-1}_{p=1}\sum^{p-1}_{q=0}
    \langle\zeta^{(i)}_{k} \zeta^{(i)}_{q}\rangle \mu^{m+p-k-q-2} (1-\beta)^{2n'+n-m-p-2}\big].
    \end{align*}
    We use the same noise correlations in \ref{noise0} (with $\xi\to\zeta$) with the addition of the correlation for the shared noise $\chi$
    \begin{equation}
    \label{noise1}
    \begin{aligned}
    \langle \chi_{m} \chi_{k}\rangle&=\sigma^2_{\chi}\delta_{mk}.
    \end{aligned}
        \end{equation}
    This simplifies the ACF to
    \begin{align*}
    A(n)&\propto \lim_{n' \to \infty}\big[(1-\beta)^{2n'+n} \langle\epsilon^{(i)}_{0}\epsilon^{(i)}_{0}\rangle+ \langle\epsilon^{(i)}_{0} y^{(i)}_{0}\rangle ( \sum^{n+n'-1}_{m=0}{\mu^m (1-\beta)^{2n'+n-m-1}} + \sum^{n'-1}_{m=0}{\mu^m (1-\beta)^{2n'+n-m-1}} )\\
    &+\langle y^{(i)}_{0} y^{(i)}_{0}\rangle\sum^{n+n'-1}_{m=0}\sum^{n'-1}_{k=0}{\mu^{m+k} (1-\beta)^{2n'+n-m-k-2} }+ \sum^{n'-1}_{k=0}{(\sigma^2_{\eta}+\sigma^2_{\chi})(1-\beta)^{2n'+n-2k-2}}\\&+ \sigma^2_{\zeta} \sum^{n'-1}_{m=1}\sum^{m-1}_{k=0}\sum^{n+n'-1}_{p=m+1}
      \mu^{m+p-2k-2} (1-\beta)^{2n'+n-m-p-2}+\sigma^2_{\zeta} \sum^{n'-1}_{m=1}\sum^{m}_{p=1}\sum^{p-1}_{q=0}
      \mu^{m+p-2q-2} (1-\beta)^{2n'+n-m-p-2}\big].
    \end{align*}
    
    \noindent We can notice that some of the sums have the form of a geometric series that has a known closed form. Using this result, the ACF becomes
    \begin{align*}
    A(n)&\propto \lim_{n' \to \infty}\big[(1-\beta)^{2n'+n} \langle\epsilon^{(i)}_{0}\epsilon^{(i)}_{0}\rangle+ \langle\epsilon^{(i)}_{0} y^{(i)}_{0}\rangle (1-\beta)^{2n'+n-1}\Big(\frac{1-(\frac{\mu}{1-\beta})^{n+n'}}{1-\frac{\mu}{1-\beta}}  + \frac{1-(\frac{\mu}{1-\beta})^{n'}}{1-\frac{\mu}{1-\beta}}\Big)\\
    &+\langle y^{(i)}_{0} y^{(i)}_{0}\rangle(1-\beta)^{2n'+n-2} \Big(\frac{1-(\frac{\mu}{1-\beta})^{n+n'}}{1-\frac{\mu}{1-\beta}}\Big)\Big( \frac{1-(\frac{\mu}{1-\beta})^{n'}}{1-\frac{\mu}{1-\beta}}\Big)+ (1-\beta)^{2n'+n-2}{(\sigma^2_{\chi} + \sigma^2_{\eta})\Big(\frac{1-\frac{1}{(1-\beta)^{2n'}}}{1-\frac{1}{(1-\beta)^2}}\Big)}\\&+ \sigma^2_{\zeta} \sum^{n'-1}_{m=1}\sum^{m-1}_{k=0}\sum^{n+n'-1}_{p=m+1}
      \mu^{m+p-2k-2} (1-\beta)^{2n'+n-m-p-2}+ \sigma^2_{\zeta} \sum^{n'-1}_{m=1}\sum^{m}_{p=1}\sum^{p-1}_{q=0}
      \mu^{m+p-2q-2} (1-\beta)^{2n'+n-m-p-2}\big].
    \end{align*}
    It can be read off from Eqs. \ref{s1} and \ref{s2} that the factors $1-\beta$ and $\mu$ must be less than 1 for the model to be stable. Otherwise, cell size and the internal component $y$ will diverge for later generations. Using this condition along with taking the limit $n' \to \infty$, the ACF simplifies to
    \begin{align*}
        A(n)&\propto(\sigma^2_{\eta}+\sigma^2_{\chi})\frac{(1-\beta)^{n}}{1-(1-\beta)^2}+\lim_{n' \to \infty} \big[ \sigma^2_{\zeta} \sum^{n'-1}_{m=1}\sum^{m-1}_{k=0}\sum^{n+n'-1}_{p=m+1}
      \mu^{m+p-2k-2} (1-\beta)^{2n'+n-m-p-2}\\&+ \sigma^2_{\zeta} \sum^{n-1}_{m=1}\sum^{m}_{p=1}\sum^{p-1}_{q=0}
      \mu^{m+p-2q-2} (1-\beta)^{2n'+n-m-p-2}\big].
    \end{align*}
    Using Mathematica to simplify the result of the remaining summations and taking the limit, the ACF has the final form 
    \begin{equation}
        A({n})\propto \sigma^2_{\chi} \frac{(1-\beta)^n}{1-(1-\beta)^2} +  \sigma^2_{\eta} \frac{(1-\beta)^n}{1-(1-\beta)^2} - \sigma^2_{\zeta} \frac{\frac{(1-\beta)^{n+1}}{1-(1-\beta)^2} + \frac{\mu^{n+1}}{\mu^2 -1}}{(\mu + \beta - 1) (1+\mu (\beta-1))},
    \end{equation}
    which can be normalized and written as
    \begin{align}
    A(n)=&\frac{1}{Z_A} \big(c_1 \ (1-\beta)^n + c_2 \ \mu^n \big),
\end{align}
with
\begin{align*}
    c_1=& \frac{ (\sigma^2_{\eta}+ \sigma^2_{\chi})}{1-(1-\beta)^2} - \frac{ \sigma^2_{\zeta} (1-\beta)}{(\mu + \beta - 1) (1+\mu (\beta-1))(1-(1-\beta)^2)},  \\
    c_2=& - \frac{ \sigma^2_{\zeta} \mu}{(\mu + \beta - 1) (1+\mu (\beta-1)) (\mu^2 -1)}, \\
    Z_A=&  c_1 + c_2.
\end{align*}
This is equivalent to the form in Eq.\ \ref{A1}.

    \subsubsection{Pearson Correlation Function}
    \noindent Substituting Eq. \ref{s3} into Eq. \ref{ps1} and noticing that the $\langle\zeta_n^{(1)} \zeta_n^{(2)}\rangle$ and $\langle\eta_n^{(1)} \eta_n^{(2)}\rangle$ noise terms collapse to zero according to Eqs. \ref{noise0}, we find
    \begin{align*}
    P(n)&\propto(1-\beta)^{2n} \langle\epsilon^{2}_{0}\rangle +2 \sum^{n-1}_{m=0} \mu^m (1-\beta)^{2n-m-1} \langle\epsilon_0 y_0\rangle+ \sum^{n-1}_{m=0} \sum^{n-1}_{r=0} \mu^{m+r} (1-\beta)^{2 n -m - r -2} \langle y_0^2 \rangle\\
    &\quad+ \sum^{n-1}_{m=0}\sum^{n-1}_{r=0}{\langle\chi_{m} \zeta_{r}\rangle(1-\beta)^{2n-m-r-2}}\\
    &=(1-\beta)^{2n} \langle\epsilon^{2}_{0}\rangle + 2 (1-\beta)^{2n-1} \Big( \frac{1-(\frac{\mu}{1-\beta})^n}{1-\frac{\mu}{1-\beta}} \Big) \langle\epsilon_0 y_0\rangle\\
    &\quad+ (1-\beta)^{2 n-2} \langle y_0^2\rangle \Big( \frac{1-(\frac{\mu}{1-\beta})^n}{1-\frac{\mu}{1-\beta}} \Big)^2+ \sigma^2_{\chi} \Big( \frac{1-(1-\beta)^{2n} }{\beta(2-\beta)} \Big)\\
    &=(1-\beta)^{2n} \langle\epsilon^{2}_{0}\rangle + 2 (1-\beta)^{n} \Big( \frac{\mu^n -(1-\beta)^n}{\beta+\mu-1} \Big) \langle\epsilon_0 y_0\rangle\\
    &\quad+ \langle y_0^2\rangle \Big( \frac{(1-\beta)^n-\mu^n}{\beta +\mu -1} \Big)^2+ \sigma^2_{\chi} \Big( \frac{1-(1-\beta)^{2n} }{\beta(2-\beta)} \Big),
    \end{align*}
    where $\epsilon_0^{(1)}$, $y_0^{(1)}$ are equal to $\epsilon_0^{(2)}$, $y_0^{(2)}$ due to the fact that sister cells share the same mother. Now the task is to find the initial moments of the system $\langle\epsilon_0^2\rangle$, $\langle y_0^2\rangle$, and $\langle\epsilon_0 y_0\rangle$. The moments are defined as
    \begin{align}
      \langle\epsilon_0^2\rangle &= \lim_{n \to \infty} \langle\epsilon_n^2\rangle,\\
      \langle y_0^2\rangle &= \lim_{n \to \infty} \langle y_n^2\rangle,\\
      \langle\epsilon_0 y_0\rangle &= \lim_{n \to \infty} \langle\epsilon_n y_n\rangle,
    \end{align}
    which can be found to be
    \begin{align}
    \label{epsilon02}
        \langle\epsilon_0^2\rangle =\ &  \frac{\sigma^2_{\chi} + \sigma^2_{\eta}}{\beta (2-\beta)} +  \sigma^2_{\zeta} \frac{\mu (\beta -1) -1}{\beta (2-\beta) (\mu^2 -1) (1+\mu (\beta -1))},\\\label{y02}
      \langle y_0^2\rangle  =\ & \frac{ \sigma^2_{\zeta}}{1-\mu^2},\\ \label{epsilon0y0}
      \langle\epsilon_0 y_0\rangle =\ & \frac{\sigma^2_{\zeta} \mu}{(1+\mu (\beta-1)) (1-\mu^2)}.
    \end{align}
    Eqs. \ref{epsilon02}, \ref{y02} and \ref{epsilon0y0} can be derived using the same methods used in the derivation of the ACF. Plugging the moments equations in the equation for the PCF and normalizing, we find the normalized PCF to be
\begin{align}
    P(n)=\frac{1}{Z_P} \big(c_0 + c_3 \ \mu^n (1-\beta)^n + c_4 \ \mu^{2n} + c_5 \ (1-\beta)^{2n} \big), 
\end{align}
        where
\begin{align*}
    c_0=& \frac{ \sigma^2_{\chi}}{\beta (2-\beta)}, \\
    c_3=& -\frac{2 \sigma^2_{\zeta}}{(\mu (\beta -1) +1) (\mu + \beta -1)^2}, \\
    c_4=& -\frac{ \sigma^2_{\zeta}}{(\mu + \beta -1)^2 (\mu^2 -1)}, \\
    c_5=& -\frac{ \sigma^2_{\zeta}}{\beta (\beta -2) (\mu + \beta -1)^2} - \frac{ \sigma^2_{\eta}}{\beta (\beta -2) }, \\
    Z_P=&  c_0 + c_3 + c_4 + c_5.
\end{align*}
This is equivalent to the form in Eq.\ \ref{P1}.
        
\section{Model with phase dependence (Eqs.\ \ref{epsilon} and \ref{delta3})}
\noindent
This model is described by two coupled equations making it impossible to obtain a closed form for cell size $\epsilon_n$ from the recurrence relations. Instead, we utilize a method that allows us to transform discrete recursion relations to algebraic equations that can be solved and inverted back to get the solution. This is the method of the $\zstroke$-transformation. In the $\zstroke$-transformation $\epsilon_{n}\rightarrow E(\zstroke)$ with $E(\zstroke)=\sum_{n=0}^{\infty}{\epsilon_n \zstroke^{-n}}$ and the inversion equation $\epsilon_n= \frac{1}{2 \pi i}   \oint\limits E(\zstroke) \zstroke^{n-1} \ \mathrm{d}\zstroke$. To apply this to the model equations we multiply Eqs. \ref{epsilon} and \ref{delta3} with $\zstroke$ and sum over $n$ from $n=0$ to $n=\infty$. The transformation is
\begin{align*}
        & \epsilon^{(i)}_{n}\rightarrow E^{(i)}(\zstroke) \ , \ \epsilon^{(i)}_{n+1}\rightarrow \zstroke (E^{(i)}(\zstroke)-\epsilon_0),\\
        & \delta^{(i)}_{n}\rightarrow \Delta^{(i)}(\zstroke) \ , \ \delta^{(i)}_{n+1}\rightarrow \zstroke (\Delta^{(i)}(\zstroke)-\delta_0),\\
        & \eta^{(i)}_{n}\rightarrow H^{(i)}(\zstroke) \ , \ \xi^{(i)}_{n}\rightarrow Z^{(i)}(\zstroke),
\end{align*}
where
\begin{align*}
        E^{(i)}(\zstroke)=&\sum_{n=0}^{\infty}{\epsilon^{(i)}_n \zstroke^{-n}}, \\
        \Delta^{(i)}(\zstroke)=&\sum_{n=0}^{\infty}{\delta^{(i)}_n \zstroke^{-n}}.
\end{align*}

\noindent Eqs. \ref{epsilon} and \ref{delta3} then become
\begin{align}
\label{Ez}
        E^{(i)}(\zstroke) (\zstroke - 1)=& \zstroke \epsilon_0 + \Delta^{(i)}(\zstroke) + H^{(i)}(\zstroke), \\ \label{deltaz}
      \Delta^{(i)}(\zstroke) (\zstroke-\lambda) =& \zstroke \delta_0 -\beta \zstroke (E^{(i)}(\zstroke) -\epsilon_0)+Z^{(i)}(\zstroke).
\end{align}
We can write Eqs. \ref{Ez} and \ref{deltaz} in matrix form as

\begin{equation}
\begin{pmatrix}
\zstroke -1 & -1  \\
\beta \zstroke & \zstroke - \lambda
\end{pmatrix}
\begin{pmatrix}
E^{(i)}(\zstroke) \\
\Delta^{(i)}(\zstroke)
\end{pmatrix}
= \begin{pmatrix}
\zstroke \epsilon_0 + H^{(i)} (\zstroke) \\
\zstroke \delta_0 + \beta \epsilon_0 \zstroke + Z^{(i)} (\zstroke)
\end{pmatrix},
\end{equation}
with solution
\begin{align}
\label{Ezsol}
        E^{(i)}(\zstroke)=& \frac{Z^{(i)}(\zstroke) + \delta_0 \zstroke + H^{(i)}(\zstroke) (\zstroke - \lambda) + \epsilon_0 \zstroke (\zstroke +\beta - \lambda)}{\zstroke (\zstroke +\beta -\lambda -1) + \lambda},
        \\
      \Delta^{(i)}(\zstroke)=& \frac{Z^{(i)}(\zstroke) (\zstroke -1) - H^{(i)}(\zstroke) \zstroke \beta + \delta_0 \zstroke (\zstroke -1) - \epsilon_0 \zstroke \beta }{\zstroke (\zstroke +\beta -\lambda -1) + \lambda}.
\end{align}

\noindent Now it is a problem of inverting Eq. \ref{Ezsol} using $\epsilon^{(i)}_n= \frac{1}{2 \pi i}   \oint\limits E^{(i)}(\zstroke) \zstroke^{n-1} \ \mathrm{d}\zstroke$. Inverting at this stage is difficult because of the noise terms $Z^{(i)}(\zstroke)$ and $H^{(i)}(\zstroke)$. Instead we carry on our calculation for the ACF and PCF and do the integral at the end to benefit from the noise properties $\langle \xi^{(i)}_{m} \xi^{(j)}_{n}\rangle=\sigma^2_{\xi}\delta_{mn}\delta_{ij}$ and $\langle\eta^{(i)}_{m} \eta^{(j)}_{n}\rangle = \sigma^2_{\eta}\delta_{mn}\delta_{ij}$.

\subsubsection{Autocorrelation Function}

\noindent The ACF is given by
\begin{align}
\label{s20}
A(n)\propto\lim_{n' \to \infty}{\langle \epsilon^{(i)}_{n+n'} \epsilon^{(i)}_{n'} \rangle} =\lim_{n' \to \infty} \frac{-1}{4 \pi^2}   \oint\limits \oint\limits \langle E^{(i)}(\zstroke_1) E^{(i)}(\zstroke_2) \rangle \zstroke_1^{n'-1} \zstroke_2^{n+n'-1} \ \mathrm{d}\zstroke_1 \mathrm{d}\zstroke_2.
\end{align}

\noindent Using Eq. \ref{Ezsol} we find
\begin{equation}
\begin{aligned}
\label{s21}
    \langle E^{(i)}(\zstroke_1) E^{(i)}(\zstroke_2) \rangle=&\frac{\langle Z^{(i)}(\zstroke_1) Z^{(i)}(\zstroke_2)  \rangle + \zstroke_1 \zstroke_2 \langle \delta_0^2 \rangle + (\zstroke_1 - \lambda) (\zstroke_2 - \lambda) \langle H^{(i)}(\zstroke_1) H^{(i)}(\zstroke_2) \rangle}{(\zstroke_1 (\zstroke_1 +\beta -\lambda -1) + \lambda)(\zstroke_2 (\zstroke_2 +\beta -\lambda -1) + \lambda)}\\ +& \frac{\zstroke_1 \zstroke_2 (\zstroke_1 + \beta -\lambda) (\zstroke_2 + \beta - \lambda) \langle \epsilon_0^2 \rangle + \zstroke_1 \zstroke_2 (\zstroke_2 + \beta - \lambda) \langle \delta_0 \epsilon_0 \rangle}{(\zstroke_1 (\zstroke_1 +\beta -\lambda -1) + \lambda)(\zstroke_2 (\zstroke_2 +\beta -\lambda -1) + \lambda)},
\end{aligned}
\end{equation}
with
\begin{equation}
\begin{aligned}
\label{s22}
        \langle Z^{(i)}(\zstroke_1) Z^{(j)}(\zstroke_2)  \rangle&= \sum_{n=0}^{\infty} \sum_{m=0}^{\infty}{ \langle \xi^{(i)}_{m} \xi^{(j)}_{n}\rangle \zstroke_1^{-n}\zstroke_2^{-m}} = \sum_{n=0}^{\infty} \sum_{m=0}^{\infty}{ \sigma^2_{\xi}\delta_{mn}\delta_{ij} \zstroke_1^{-n}\zstroke_2^{-m}}= \frac{ \sigma^2_{\xi} \delta_{ij}}{1-\frac{1}{\zstroke_1 \zstroke_2}}, \\
        \langle H^{(i)}(\zstroke_1) H^{(j)}(\zstroke_2)  \rangle&= \sum_{n=0}^{\infty} \sum_{m=0}^{\infty}{ \langle \eta^{(i)}_{m} \eta^{(j)}_{n}\rangle \zstroke_1^{-n}\zstroke_2^{-m}} = \sum_{n=0}^{\infty} \sum_{m=0}^{\infty}{ \sigma^2_{\eta}\delta_{mn}\delta_{ij} \zstroke_1^{-n}\zstroke_2^{-m}}=\frac{\sigma^2_{\eta} \delta_{ij}}{1-\frac{1}{\zstroke_1 \zstroke_2}}.\\
\end{aligned}
\end{equation}
Substituting Eqs. \ref{s22} in \ref{s21}, then \ref{s21} in \ref{s20}, the ACF can be found applying the residue theorem to be
\begin{align}
    A(n)=\frac{1}{Z_A} \big(q_{-} a_{-}^n + q_{+} a_{+}^n \big),
\end{align}
as in Eq.\ \ref{A2}, where
\begin{align}\label{qpm}
    a_\pm=& \frac{\lambda - \beta +1 \pm \sqrt{(\beta-\lambda-1)^2 - 4 \lambda}}{2} \ , \ \ |a_\pm|<1  \ (\text{Stability Condition})\\
    \label{qpm2}
    q_\pm =& \pm\frac{[\lambda(1+a_\pm^2)-a_\pm(1+\lambda^2)]\sigma^2_\eta-a_\pm\sigma^2_\xi}{(a_\pm^2-1)},\\
    Z_A=&  q_{-} + q_{+}.
\end{align}
Given that, as defined in the main text,
\begin{equation}
\label{ro}
r = (1+\beta-\lambda)/2, \qquad \omega^2 = \beta-r^2,
\end{equation}
Eq.\ \ref{qpm} becomes $a_\pm = 1-r\pm\sqrt{-\omega^2}$. Furthermore, one can verify from Eq.\ \ref{qpm} that
\begin{equation}
\label{lba}
\lambda = a_+a_-, \qquad \beta = (a_+-1)(a_--1).
\end{equation}
Thus, because $\beta$ and $\lambda$ are functions of $a_\pm$, it is clear that $q_\pm$ is a function of $a_\pm$ and the noise strengths $\sigma_\eta^2$ and $\sigma_\xi^2$ as given by Eq.\ \ref{qpm2}.

\subsubsection{Pearson Correlation Function}
\noindent Similarly the PCF is
    \begin{align}
    P(n)\propto\langle\epsilon^{(1)}_n\epsilon^{(2)}_n\rangle=\frac{-1}{4 \pi^2}   \oint\limits \oint\limits \langle E^{(1)}(\zstroke_1) E^{(2)}(\zstroke_2) \rangle \zstroke_1^{n-1} \zstroke_2^{n-1} \ \mathrm{d}\zstroke_1 \mathrm{d}\zstroke_2,
    \end{align}
    and
\begin{align}
    \langle E^{(1)}(\zstroke_1) E^{(2)}(\zstroke_2) \rangle=&\frac{\langle Z^{(1)}(\zstroke_1) Z^{(2)}(\zstroke_2)  \rangle + \zstroke_1 \zstroke_2 \langle \delta_0^2 \rangle + (\zstroke_1 - \lambda) (\zstroke_2 - \lambda) \langle H^{(1)}(\zstroke_1) H^{(2)}(\zstroke_2) \rangle}{(\zstroke_1 (\zstroke_1 +\beta -\lambda -1) + \lambda)(\zstroke_2 (\zstroke_2 +\beta -\lambda -1) + \lambda)} \nonumber\\
    &+ \frac{\zstroke_1 \zstroke_2 (\zstroke_1 + \beta -\lambda) (\zstroke_2 + \beta - \lambda) \langle \epsilon_0^2 \rangle + \zstroke_1 \zstroke_2 (\zstroke_2 + \beta - \lambda) \langle \delta_0 \epsilon_0 \rangle}{(\zstroke_1 (\zstroke_1 +\beta -\lambda -1) + \lambda)(\zstroke_2 (\zstroke_2 +\beta -\lambda -1) + \lambda)}\nonumber\\
    =&\frac{\zstroke_1 \zstroke_2 \langle \delta_0^2 \rangle +\zstroke_1 \zstroke_2 (\zstroke_1 + \beta -\lambda) (\zstroke_2 + \beta - \lambda) \langle \epsilon_0^2 \rangle + \zstroke_1 \zstroke_2 (\zstroke_2 + \beta - \lambda) \langle \delta_0 \epsilon_0 \rangle }{(\zstroke_1 (\zstroke_1 +\beta -\lambda -1) + \lambda)(\zstroke_2 (\zstroke_2 +\beta -\lambda -1) + \lambda)}.
\end{align}
Using residue theorem we find
\begin{equation}
\label{pn2}
\begin{aligned}
    P(n)\propto& \Big(\frac{a_{-}^n-a_{+}^n}{a_{-}-a_{+}}\Big)^2 \langle \delta_0^2 \rangle + \Big(\frac{a_{-}^n (a_{-} +\beta -\lambda)-a_{+}^n (a_{+} +\beta -\lambda)}{a_{-}-a_{+}}\Big)^2 \langle \epsilon_0^2 \rangle \\ 
    &+ 2 \Big(\frac{a_{-}^n-a_{+}^n}{a_{-}-a_{+}}\Big) \Big(\frac{a_{-}^n (a_{-} + \beta - \lambda)-a_{+}^n (a_{+}+\beta - \lambda)}{a_{-}-a_{+}}\Big) \langle \delta_0 \epsilon_0 \rangle,
\end{aligned}
\end{equation}
with moments
\begin{equation}
\label{s27}
    \begin{aligned}
      \langle\epsilon_0^2\rangle &= \lim_{n \to \infty} \langle\epsilon_n^2\rangle=\frac{\sigma^2_{\xi} (1+a_{+}a_{-}) + \sigma^2_{\eta} (\lambda^2 (a_{-}a_{+}+1)-2\lambda (a_{-}+a_{+}) + a_{-} a_{+}+1)}{(a^2_{-}-1)(a^2_{+}-1)(1-a_{-}a_{+})},\\
      \langle \delta_0^2\rangle &= \lim_{n \to \infty} \langle \delta_n^2\rangle=\frac{2 \sigma^2_{\xi} (a_{+}-1)(a_{-}-1) + \sigma^2_{\eta} \beta^2 (a_{+}a_{-}+1) }{(a^2_{-}-1)(a^2_{+}-1)(1-a_{-}a_{+})},\\
      \langle\epsilon_0 \delta_0\rangle &= \lim_{n \to \infty} \langle\epsilon_n \delta_n\rangle=\frac{\sigma^2_{\xi} (a_{+}-1)(a_{-}-1) + \sigma^2_{\eta} \beta (-\lambda (a_{+}+a_{-})+a_{+}a_{-}+1)}{(a^2_{-}-1)(a^2_{+}-1)(a_{-}a_{+}-1)}.
    \end{aligned}
    \end{equation}
    Substituting Eqs. \ref{s27} into \ref{pn2} the PCF can be written as
    \begin{align}
    P(n)=&\frac{1}{Z_P} \big(c_3 \ a_{-}^{2n} + c_4 \ a_{+}^{2n} + c_5 \ a_{-}^{n} a_{+}^{n}\big),
    \end{align}
as in Eq.\ \ref{P2},    where
    \begin{align*}
    c_3=& \frac{\langle \delta_0^2\rangle + 2 \langle \delta_0 \epsilon_0\rangle (a_{-} + \beta - \lambda) + \langle \epsilon_0^2\rangle (a_{-} + \beta - \lambda)^2}{(a_{-}-a_{+})^2},\\
    c_4=& \frac{\langle \delta_0^2\rangle + 2 \langle \delta_0 \epsilon_0\rangle (a_{+} + \beta - \lambda) + \langle \epsilon_0^2\rangle (a_{+} + \beta - \lambda)^2}{(a_{-}-a_{+})^2}, \\
    c_5=& \frac{-2 \langle \delta_0^2\rangle - 2 \langle\epsilon_0 \delta_0\rangle (a_{-} + a_{+} + 2 (\beta - \lambda)) - 2 \langle\epsilon_0^2\rangle (a_{+} (\beta-\lambda) + (\beta - \lambda)^2 + a_{-} (a_{+} + \beta - \lambda)) }{(a_{-}-a_{+})^2}, \\
    Z_P=& \langle \epsilon_0^2 \rangle = c_3 + c_4 + c_5.
\end{align*}
From Eqs.\ \ref{lba} and \ref{s27} it is clear that $c_3$, $c_4$, and $c_5$, which are called $s_-$, $s_+$, and $s$ in the main text, are functions of $a_\pm$ and the noise strengths $\sigma_\eta^2$ and $\sigma_\xi^2$.

\subsubsection{Stability}
\noindent To determine the range of parameters over which the model is stable we can write the model in terms of a transition matrix

\begin{equation}
\label{transmatrix}
    \begin{pmatrix} 
	\epsilon_{n+1}^{(i)} \\
	\delta_{n+1}^{(i)} \\
	\end{pmatrix}=\begin{pmatrix} 
	1 & 1 \\
	-\beta & \lambda-\beta \\
	\end{pmatrix}
	\begin{pmatrix} 
	\epsilon_n^{(i)} \\
	\delta_n^{(i)} \\
	\end{pmatrix}.
\end{equation}
Iterating Eq. \ref{transmatrix} gives
\begin{equation}
\label{transmatrix2}
    \begin{pmatrix} 
	\epsilon_{n+1}^{(i)} \\
	\delta_{n+1}^{(i)} \\
	\end{pmatrix}=\begin{pmatrix} 
	1 & 1 \\
	-\beta & \lambda-\beta \\
	\end{pmatrix}^n
	\begin{pmatrix} 
	\epsilon_0^{(i)} \\
	\delta_0^{(i)} \\
	\end{pmatrix}.
\end{equation}
We write Eq. \ref{transmatrix2} as
\begin{equation}
    X(n+1)=A^n X(0).
\end{equation}
Defining $D^n=PA^nP^{-1}$ and $Y(n)=PX(n)$, we get
\begin{equation}
    Y(n+1)=D^nY(0),
\end{equation}
where $D$ is a diagonal matrix
\begin{equation}
    D=\begin{pmatrix} 
	a_{+} & 0 \\
	0 & a_{-} \\
	\end{pmatrix},
\end{equation}
with $a_\pm=(\lambda-\beta+1\pm\sqrt{(\beta-\lambda-1)^2-4\lambda})/2$ the eigenvalues of $A$. Thus $    D^n=\begin{pmatrix} 
	a_{+}^n & 0 \\
	0 & a_{-}^n \\
	\end{pmatrix} $. In the limit $n\to \infty$ the system should approach steady state such that
	\begin{equation}
	D Y = Y
	\end{equation}
	which has the solution $Y=0$. This can only be achieved if $\abs{a_{\pm}}<1$. Using the definitions in Eq.\ \ref{ro}, $a_\pm$ becomes
\begin{equation}
\label{apm}
a_\pm = 1-r\pm\sqrt{-\omega^2}.
\end{equation}
For the case $\omega^2>0$, the stability condition $|a_\pm|<1$ implies $(1-r)^2+\omega^2<1$, or 
\begin{equation}
\label{cond1}
\omega^2 < -r(r-2).
\end{equation}
For the case $\omega^2<0$, $a_\pm$ is real, and the stability condition reads $a_+<1$ and $-a_-<1$. The first condition implies $1-r+\sqrt{-\omega^2} < 1$, or
\begin{equation}
\label{cond2}
\omega^2 > -r^2.
\end{equation}
The second condition implies $-(1-r-\sqrt{-\omega^2}) < 1$, or
\begin{equation}
\label{cond3}
\omega^2 > -(r-2)^2.
\end{equation}
The conditions in Eqs. \ref{cond1}, \ref{cond2} and \ref{cond3} define the outer boundaries in Fig.\ \ref{fig3}c.

\subsubsection{Dynamic regimes}        
Oscillation occurs when Eq.\ \ref{apm} becomes complex, or
\begin{equation}
\omega^2 > 0.
\end{equation}
This condition defines the oscillatory regime in Fig.\ \ref{fig3}c. Alternation occurs when $a_-$ is negative and its magnitude is larger than that of $a_+$. These properties endow the dominant term of the ACF with a factor of $(-1)^n$, which causes it to alternate sign each generation. We will first assume that $a_-$ is negative, then verify this assumption {\it post hoc}. The magnitude condition in this case reads $-a_- > a_+$, or $-(1-r-\sqrt{-\omega^2}) > 1-r+\sqrt{-\omega^2}$, which simplifies to
\begin{equation}
\label{alt1}
r>1.
\end{equation}
We now verify that $a_-$ is negative for $r>1$. Solving the definition $a_- = 1-r-\sqrt{-\omega^2}$ for $r$, the condition $r>1$ reads $1-a_--\sqrt{-\omega^2}>1$, or $a_-<-\sqrt{-\omega^2}$. When
\begin{equation}
\label{alt2}
\omega^2<0,
\end{equation}
we see that $a_-$ is less than a real, negative quantity, and thus $a_-$ is negative. Therefore, Eqs.\ \ref{alt1} and \ref{alt2} define the alternating regime in Fig.\ \ref{fig3}c.

\section{Inferring parameter values from the data}

In all fits, the two sister lineages were fit simultaneously to obtain a single set of parameters $(\beta, \lambda)$ that capture the shared dynamics of both lineages. The dynamics fitting to data was conducted using linear regression analysis in Matlab, utilizing built-in functions. Standard error (SE) of the fit was obtained within Matlab. The ACF fit of sister lineages to data was conducted using non-linear fit functions in Matlab. An approximation of SE of the fit was obtained from the covariance matrix of the fit parameters. The covariance matrix ($\boldsymbol{C}$) can be obtained from the Jacobian matrix of the residuals ($\boldsymbol{J}$) as follows 
\begin{align}
\boldsymbol{H}=\boldsymbol{J^T}\boldsymbol{J},\\
\label{covar}\boldsymbol{C}=\boldsymbol{H^{-1}},
\end{align}
where $\boldsymbol{H}$ is the hessian matrix, $J_{nj}=\frac{\partial r_n}{\partial a_j}$, $r_n=\tilde{A}_n-A(n;\boldsymbol{a})$ is the residual of the ACF fit at the $n^{th}$ generation, $\boldsymbol{\tilde{A}}$ is the vector of measured ACF, and $\boldsymbol{a}$ are the model parameters $\beta$ and $\lambda$. SE in each parameter is then $SE(a_i)=\sqrt{C_{ii} \times MSE}$, where $MSE$ is the mean squared error of the fit. In what follows we prove this relation.

We start by linearizing the ACF and calculating the residuals of the fit as follows
\begin{align}
\boldsymbol{A}&=\boldsymbol{X}\boldsymbol{a},\\
\boldsymbol{r}&=\boldsymbol{\tilde{A}}-\boldsymbol{A}=\boldsymbol{\tilde{A}}-\boldsymbol{Xa},
\end{align}
where $\boldsymbol{X}$ is a generic matrix of time functions. It is important to notice that $\boldsymbol{X}$ is related to the Jacobian by $X_{nj}=-\frac{\partial r_n}{\partial a_j}=-J_{nj}$, making the covariance matrix $\boldsymbol{C=X^T X}$. To find the best fit of the parameters, we minimize the sum of squares of the residuals
\begin{equation}
\label{sos}
S=\sum_{n=1}^{N} r_n^2,
\end{equation}
where $N$ is the number of generations in a lineage. This sum can be expanded as
\begin{align*}
S&=\sum_{n=1}^{N} (\tilde{A}_n-\sum_{i=1}^{M} X_{ni}a_i)(\tilde{A}_n-\sum_{j=1}^{M} X_{nj}a_j)\\
&=\sum_{n=1}^{N} (\tilde{A}_n^2 - 2 \tilde{A}_n \sum_{j=1}^{M} X_{nj}a_j + \sum_{i,j=1}^{M} X_{ni}a_i X_{nj}a_j ).
\end{align*}
We then take the derivative of $S$ with respect to the parameters
\begin{align*}
\label{sos_2}
\frac{S}{a_k}=0=\sum_{n=1}^{N} (- 2 \tilde{A}_n \sum_{j=1}^{M} X_{nj}\delta_{jk} + \sum_{i,j=1}^{M} X_{ni}X_{nj} (a_i\delta_{jk}+\delta_{ik}a_j) ),
\end{align*}
where $\delta$ is the Kronecker delta. Further simplification results in
\begin{equation}
\label{parameters_1}
\sum_{n=1}^{N} \sum_{i=1}^{M} X_{ni} X_{nk} a_i=\sum_{n=1}^{N} \tilde{A}_n X_{nk},
\end{equation}
which is the matrix equation
\begin{equation}
\label{parameters_2}
\boldsymbol{X^T}\boldsymbol{X}\boldsymbol{a}=\boldsymbol{X^T}\boldsymbol{\tilde{A}}.
\end{equation}
This can be very easily solved for the parameters $\boldsymbol{a}$
\begin{equation}
\label{parameters_3}
\boldsymbol{a}=\boldsymbol{(X^T X)^{-1}}\boldsymbol{X^T}\boldsymbol{\tilde{A}}=\boldsymbol{C}\boldsymbol{X^T}\boldsymbol{\tilde{A}}.
\end{equation}
The error in each parameter can be found through the relation
\begin{equation}
\label{error}
\boldsymbol{\delta a}=\boldsymbol{C}\boldsymbol{X^T}\boldsymbol{\delta\tilde{A}}.
\end{equation}
This illustrates that the error in each parameter is affected by the error in the data. The covariance of $\boldsymbol{a}$ is then
\begin{equation}
\label{error_2}
\sigma^2_{\boldsymbol{a}}=\boldsymbol{\langle \delta a  \delta a^T \rangle}= \boldsymbol{\langle \boldsymbol{C}\boldsymbol{X^T}\boldsymbol{\delta\tilde{A}}  (\boldsymbol{C}\boldsymbol{X^T}\boldsymbol{\delta\tilde{A}})^T \rangle}=\boldsymbol{ C X^T \langle \delta\tilde{A}  \delta\tilde{A}^T \rangle X C^T }.
\end{equation}
We assume that the data is statistically uncorrelated making $\langle \delta\tilde{A}  \delta\tilde{A}^T \rangle_{ij}=\sigma^2_i \delta_{ij}$. The variance of the data can be approximated by the mean squared error of the fit which we invoke here making $\langle \delta\tilde{A}  \delta\tilde{A}^T \rangle_{ij}=MSE \times \delta_{ij}$, therefore eq.\ \ref{error_2} becomes
\begin{equation}
\label{error_3}
\sigma^2_{\boldsymbol{a}}=\boldsymbol{C}\times MSE,
\end{equation}
which means that SE for parameter $i$ is $SE(a_i)={\sigma_{\boldsymbol{a}}}_{ii}=\sqrt{C_{ii}\times MSE}$.

Once we have the SE in parameters $\boldsymbol{a}$, we are able to invoke the relation $\sigma^2_{b_i}=\sum_{j=1}^{M} \sigma^2_{a_j} (\frac{\partial b_i}{\partial a_j})^2$ to find SE in $\boldsymbol{b}$, assuming that the covariance of parameters is zero. Using the definitions in Eqs.\ \ref{ro}, we find the SE in $r$ and $\omega^2$ to be
\begin{align}
SE(r)&=\frac{1}{2}\sqrt{ SE(\beta)^2 + SE(\lambda)^2},\\
SE(\omega^2)&=\frac{1}{\sqrt{2}}\sqrt{ (1-\beta+\lambda)^2 SE(\beta)^2 + (1+\beta-\lambda)^2 SE(\lambda)^2}.
\end{align}

\section{Supplementary Figures}

\begin{figure}[H]
\centering
\includegraphics[width=0.4\linewidth]{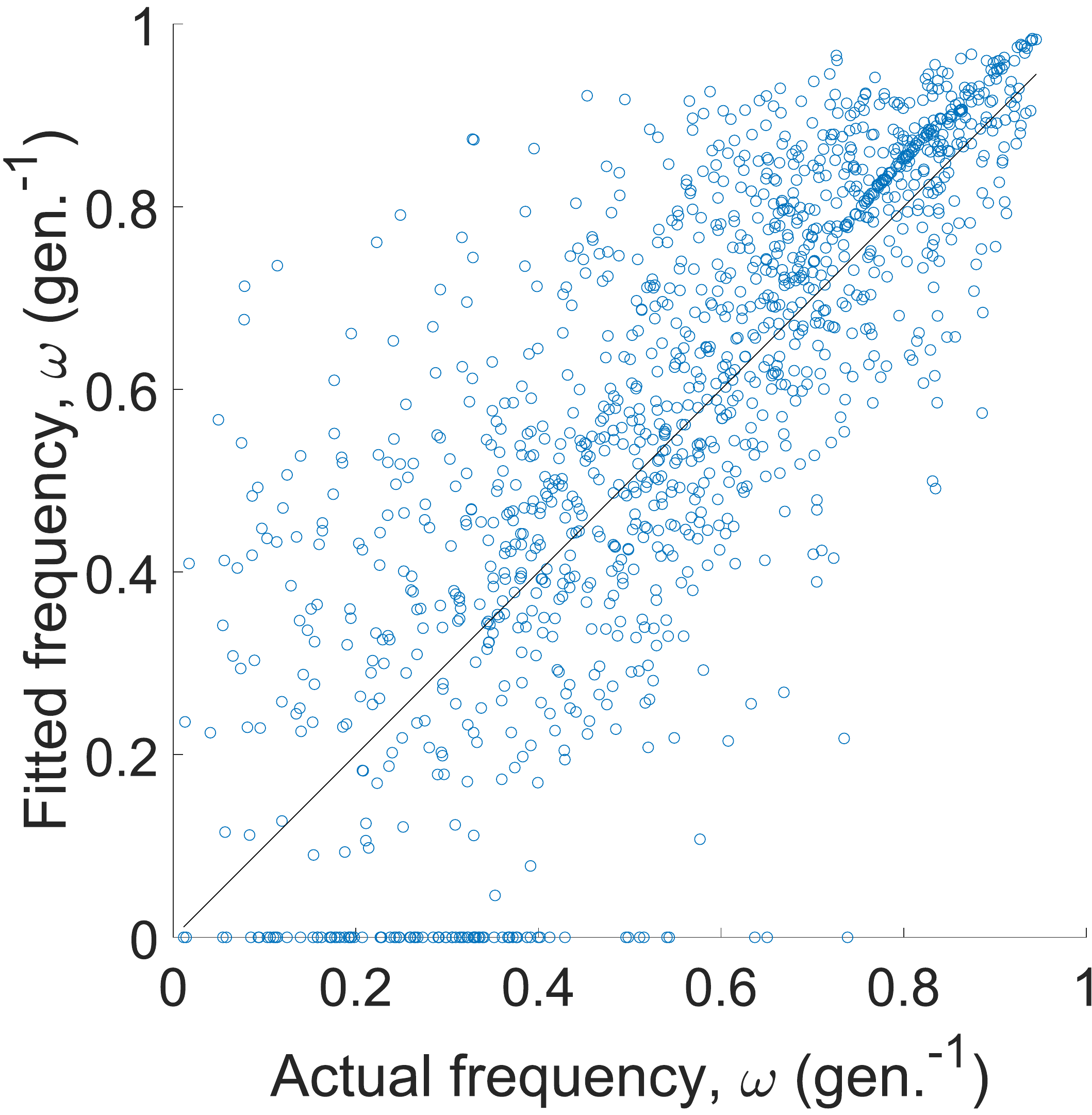}
\caption{Frequency fit from the ACF with $N=20$ generations vs.\ actual frequency [$\omega = \sqrt{\beta-r^2}$ with $r = (\lambda -\beta+1)/2$] for sampled frequencies within our model's oscillatory regime. Approximately $90\%$ of fitted frequencies correlate with the actual frequency, while the remainder are false negatives (fitted $\omega=0$).}
\end{figure}

\begin{figure}[H]
\centering
\includegraphics[width=0.9\linewidth]{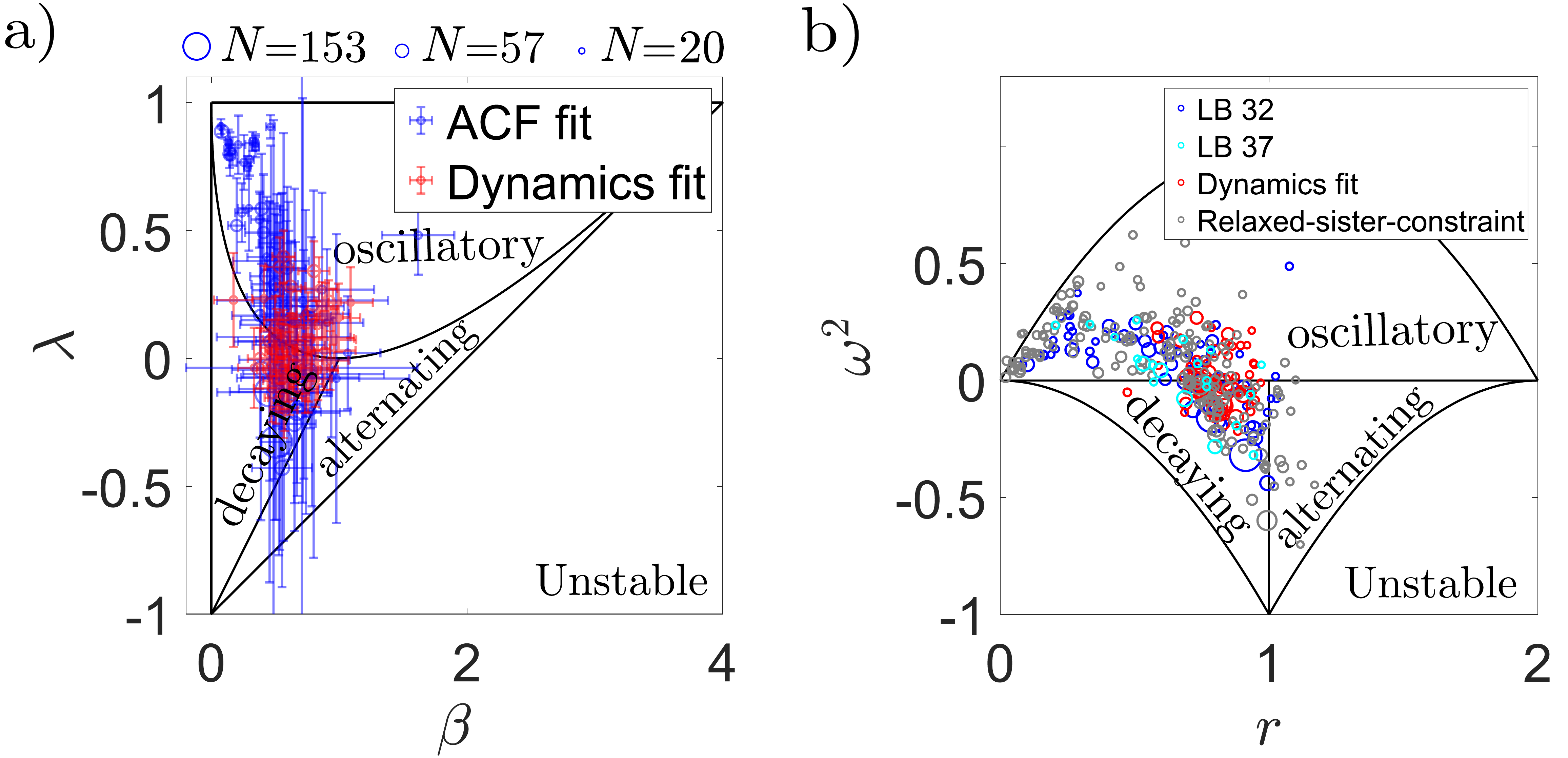}
\caption{(a) Phase space of dynamic regimes from Fig.\ \ref{fig3}c and data from Fig.\ \ref{fig4}a, here plotted in the space of $\beta$ and $\lambda$. (b) Red, blue: Data from Fig.\ \ref{fig4}a without error bars to see circle sizes. Cyan: Fits of Eq.\ \ref{A2} to ACF data in a different growth condition, LB at 37$^\circ$ (all other data is LB at 32$^\circ$). Gray: Fits of Eq.\ \ref{A2} to ACF data without the condition that sister lineages have the same parameter values.}
\end{figure}

\begin{figure}[H]
\centering
\includegraphics[width=0.9\linewidth]{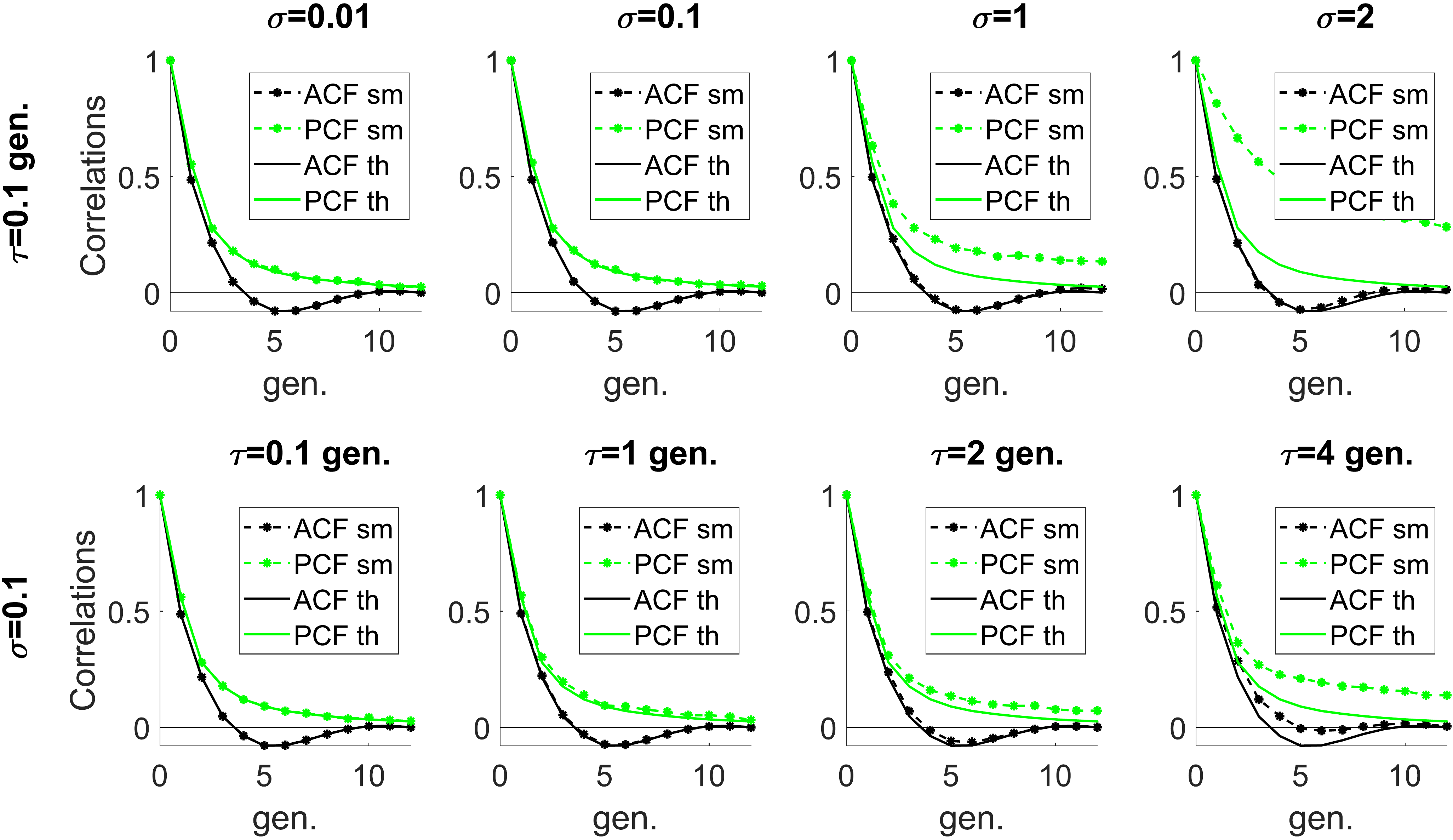}
\caption{Ornstein–Uhlenbeck simulations were performed to test the robustness of our results to time varying parameters.  Each parameter, $\beta$ and $\lambda$ (here given by the ACF fits), fluctuates around its mean according to an Ornstein-Uhlenbeck process with the specified timescale $\tau$ and standard deviation $\sigma$. We see that the correlation functions are robust to these parameter fluctuations: the theoretical curves are recovered in the limit of weak, fast noise, as expected (left); and the central observation that the PCF has a longer timescale than the ACF remains robust, even for $\tau$ as large as $4$ generations or $\sigma$ as large as $2$ (right).}
\end{figure}

\end{document}